\documentclass[12pt]{article}

\usepackage{amsmath,amssymb,amsfonts}
\usepackage{amsthm}
\usepackage{bm}
\usepackage{geometry}
\usepackage[hidelinks]{hyperref}
\usepackage{enumitem}
\usepackage{listings}
\usepackage{graphicx}
\usepackage{float}

\usepackage{algorithm}
\usepackage{algpseudocode}

\newtheorem{lemma}{Lemma}[section]

\newtheorem{definition}{Definition}[section]
\newtheorem{remark}{Remark}[section]
\newtheorem{example}{Example}[section]

\usepackage{xcolor}

\title{\textbf{Symplectic $\mathcal{H}_2$ Model Reduction for High-Dimensional 
		Linear Quantum Systems}}

\author{
	Alfio Borzì{\footnote{\textsc{A. Borzì}:
			Institut f\"ur Mathematik,
			Universit\"at W\"urzburg,
			Emil-Fischer-Strasse 30,
			97074 W\"urzburg,
			Germany. (alfio.borzi@uni-wuerzburg.de)}}
	\and 
	Guofeng Zhang{\footnote{\textsc{G. Zhang}:
			Department of Applied Mathematics, 
			The Hong Kong Polytechnic University, Hung Hom, Kowloon, Hong Kong Special
			Administrative Region, China (guofeng.zhang@polyu.edu.hk)}}
}
\date{\today}

\begin{document}
	\maketitle

\begin{abstract}
	The $\mathcal{H}_2$ model reduction problem for high-dimensional linear quantum
	systems is studied under the constraint of physical realizability (PR). This
	constraint requires preservation of the canonical commutation relations and the
	quantum input-output structure, and therefore prevents the direct use of
	standard projection methods. A symplectic Petrov-Galerkin framework is presented,
	in which reduced-order models automatically  satisfy the PR identities by construction. \\
	Within this framework, a symplectic variant of the iterative rational Krylov
	algorithm is developed and referred to as Quantum IRKA (Q-IRKA). At each
	iteration, an enriched tangential rational Krylov pool is generated from shifted
	linear solves. A symplectic basis is then extracted by a Gram-Schmidt-type
	procedure, paired with symplectic conjugates, and normalized so that the reduced
	trial space satisfies the canonical symplectic constraint. The interpolation
	points are updated from selected mirror images of the poles of the current
	reduced-order model, while the reduced-order matrices are obtained exclusively by
	structure-preserving projection. \\
	Numerical experiments on low-channel oscillator-chain systems and on a bosonic
	Kitaev-chain-inspired benchmark show that Q-IRKA is effective for large-scale
	linear quantum systems. Symplecticity and PR are preserved to machine precision,
	and accurate reduced-order models are obtained with moderate computational cost. The
	results also show that reduction quality depends substantially on dissipation
	geometry, channel placement, heterogeneity, and reduced order. These findings
	indicate that scalable $\mathcal{H}_2$ model reduction of linear quantum systems
	can be achieved while strictly preserving the underlying physical structure.
\end{abstract}
	

	\section{Introduction}

	Projection-based $\mathcal{H}_2$ model reduction is a fundamental tool in
	systems and control theory and scientific computing; see, e.g.,
	\cite{YL99,GugercinAntoulas2004,Antoulas2005,GugercinAntoulasBeattie2008,BennerGugercinWillcox2015,ABG20,JB25,ZL02}.
	For linear quantum systems, however, model reduction is constrained by
	physical realizability (PR), which ensures compatibility with the
	Heisenberg evolution of open quantum harmonic oscillators
	\cite{JamesNurdinPetersen2008,Petersen2016,ZhangDong2022}.
	In quadrature coordinates, PR appears as a set of algebraic identities
	coupling system matrices $(A,B,C,D)$ through the canonical symplectic matrices.

	Physical realizability constraints fundamentally alter the model reduction problem for linear quantum systems. While
	classical $\mathcal{H}_2$ model reduction can be formulated as an unconstrained
	optimization problem (see, e.g., \cite{ZDG96, YL99,ZL02,JB25}), the quantum case requires preservation of nonlinear
	PR identities at the reduced level. Direct enforcement of these constraints
	leads to nonlinear algebraic conditions that are difficult to reconcile
	with scalable large-scale algorithms \cite{WuXueZhangPetersen2025}. This motivates reduction strategies
	that preserve PR by construction.

	Reduced PR models are particularly relevant in repeated workflows where
	full-order simulations become expensive, such as design and optimization loops,
	controller and filter synthesis, and scalability studies over varying network
	size and channel structure. In such settings, one needs reduced-order models that
	remain physically consistent, and computable using scalable sparse
	linear-algebra operations.

	The literature on model reduction of quantum systems is extensive and may be broadly categorized according to the underlying dynamical representation.
	Within the Heisenberg picture, singular perturbation techniques have been applied to reduce passive linear quantum systems \cite{Petersen12},  classical balanced truncation methods have been successfully generalized to the linear quantum domain \cite{Nurdin14}, a tangential interpolatory
	projection framework  has been developed in \cite{TechakesariNurdin2017}, and $\mathcal{H}_2$ optimal model reduction for quantum linear systems have been proposed in \cite{WuXueZhangPetersen2025}. Conversely, in the Schr\"{o}dinger picture, the development of projection filters for stochastic master equations (SMEs) \cite{HandelandMabuchi2005,GZP19,GZP20,GDPD20,RAM25} has facilitated the design of measurement feedback stabilizing controllers \cite{AMR25,LD25}. More recently, an operator projection technique was introduced to identify minimal  realizations for discrete-time quantum walks by constructing effective subspaces from reachable sets or observables \cite{ticozzi2023}. This framework has since been expanded to encompass continuous-time Markovian master equations \cite{ticozzi2024b}, discrete-time conditional systems with generalized measurements \cite{ticozzi2024}, and continuous-time quantum filters \cite{ticozzi2025}.

	In this work, we adopt the symplectic Petrov-Galerkin
	projection. The central observation is that if the trial basis
	\(V\in\mathbb{R}^{2n\times 2r}\) satisfies
	\[
	V^\top J_n V = J_r,
	\]
	and if the test basis is chosen as
	\[
	U = J_r^{-1}V^\top J_n,
	\]
	where the canonical symplectic matrix
	\[
	J_k := I_k \otimes
	\begin{bmatrix}
		0 & 1\\
		-1 & 0
	\end{bmatrix}
	= I_k\otimes J_2,
	\]
	then projection of a PR full-order model automatically yields a reduced-order model
	that remains PR. The model reduction problem is thus recast as the construction
	of effective symplectic approximation spaces rather than the direct solution of
	a \emph{constrained} reduced-order optimization problem (see for example \cite{WuXueZhangPetersen2025}).

	Within this framework, we develop a symplectic variant of the iterative
	rational Krylov algorithm (IRKA), referred to as Quantum IRKA (Q-IRKA).
	The method combines tangential rational interpolation with symplectic
	basis extraction and normalization, and constructs reduced-order models
	exclusively through structure-preserving projection. At each iteration, the interpolation points are updated from the poles of the
	current reduced-order model, and a new reduced-order system is obtained exclusively through
	symplectic projection. In particular, the reduced matrices are not modified
	independently of the projection space. As a consequence,
	physical realizability is preserved to machine precision throughout
	the iteration.

	Our approach is closely related to the PR-preserving tangential interpolatory
	projection framework developed in \cite{TechakesariNurdin2017}. In that work,
	the reduced-order model is constructed by a structure-preserving interpolatory
	projection that matches prescribed tangential interpolation conditions while
	retaining physical realizability. The interpolation points and tangential
	directions are selected as external design data (possibly guided by auxiliary
	selection criteria), and the resulting reduction is not posed as a \emph{recursive}
	pole-updating iteration.

	The present work differs in that the tangential interpolatory construction is
	embedded into an IRKA-type fixed-point iteration. More precisely, the
	interpolation points are updated recursively from the poles of the current
	reduced-order model, and after each shift update a new PR-preserving reduced-order system
	is recomputed by symplectic projection. In this way, Q-IRKA combines the
	structure-preserving quantum interpolatory framework of
	\cite{TechakesariNurdin2017} with the recursive pole-mirroring update of
	the classical \(\mathcal H_2\) IRKA \cite{GugercinAntoulasBeattie2008}.

	This distinction is
	particularly important in large-scale settings, where direct \(\mathcal H_2\)
	or \(\mathcal H_\infty\) optimization formulations for the selection of interpolation points (as adopted in \cite{TechakesariNurdin2017}) can be computationally
	demanding; see, for example, the recent discussion in \cite{JB25}.  By contrast, Q-IRKA is built around repeated shifted linear solves and
	low-dimensional reduced-order updates, which makes it attractive from a
	computational viewpoint. Consequently, Q-IRKA should be viewed not simply as another interpolatory
	projection method, but as a structure-preserving recursive interpolatory
	reduction algorithm for linear quantum systems.

	A second objective of this paper is to understand how reduction quality depends
	on physically meaningful structural features of the underlying quantum network.
	To this end, we study  low-channel port-Hamiltonian
	configurations \cite{PolyugavanDerSchaft2010}, and a bosonic Kitaev-chain-inspired benchmark \cite{MPC18,BQND24} motivated by the recent
	optomechanical realization reported in \cite{Slim2024}. These models allow us
	to examine how dissipation geometry, channel aggregation, and problem size
	influence stability margins, Hankel singular value decay, and attainable
	\(\mathcal{H}_2\) accuracy.

	In summary, the main contributions of this work are twofold.
	First, we introduce Q-IRKA, a scalable structure-preserving recursive
	$\mathcal{H}_2$ reduction method for linear quantum systems, in which
	physical realizability is enforced exactly by symplectic projection.
	Second, we provide a systematic numerical study showing that reduction
	performance depends not only on the system dimension, but also on the
	geometry of dissipation and on the number and distribution of external
	channels. In particular, large-scale experiments indicate that scalability
	must be assessed jointly in terms of computational cost, structural
	robustness, and approximation quality.
	
	For simplicity of presentation, the quantum linear systems considered here are square. The proposed symplectic projection framework and recursive interpolatory construction extend in a straightforward way to non-square linear quantum systems with distinct input and output dimensions.

	The paper is organized as follows.
	Section~\ref{sec-full-order-transfer} introduces the class of full-order
	linear quantum systems considered in this work, together with the associated
	transfer matrix and the physical realizability identities.
	Section~\ref{sec-H2-model-red} formulates the \(\mathcal{H}_2\) model reduction
	problem and presents the symplectic projection framework used to preserve
	physical realizability by construction.
	Section~\ref{sec:symp_qirka} develops the Q-IRKA algorithm and discusses its
	practical implementation details.
	The subsequent two sections present the benchmark families and the corresponding
	numerical results, including low-channel port-Hamiltonian systems and
	bosonic Kitaev-chain-inspired models.
	A section of conclusion completes this work.

	\textbf{Notation.} $i=\sqrt{-1}$.   \(\Im(\lambda)\) and  \(\Re(\lambda)\)  are the imaginary part and real  part of the complex number \(\lambda \) respectively.  \(\mathrm{spec}(A)\)  is the set of eigenvalues of the matrix $A$.  $*$ denotes conjugate transpose of a vector or matrix. $R \succeq 0$ means that the matrix \(R\) is positive semi-definite.  $\operatorname{Range}(R)$ denotes  the image space generated by the columns of the matrix $R$.

	\section{Full-order system and transfer matrix}
	\label{sec-full-order-transfer}

	Consider a Hurwitz stable linear quantum system
	\begin{equation}
		\dot x = A x + B w,
		\qquad
		y = C x + D w,
		\label{eQuantumSystem}
	\end{equation}
	where \(x(t)\) is a column vector of \(2n\) self-adjoint system operators
	(canonical variables), \(w(t)\) is a column vector of \(2m\) input field
	quadratures, and \(y(t)\) is a column vector of \(2m\) output field
	quadratures. The associated system matrices are
	\[
	A\in\mathbb{R}^{2n\times 2n},
	B\in\mathbb{R}^{2n\times 2m},
	C\in\mathbb{R}^{2m\times 2n},
	D\in\mathbb{R}^{2m\times 2m}.
	\]
	The dimension \(2n\) reflects that each bosonic mode contributes one pair of
	canonical coordinates.

	The input-output transfer matrix  associated  with the system \eqref{eQuantumSystem} is
	\[
	\Xi_G(s):=C(sI-A)^{-1}B+D.
	\]

	Since the feed-through matrix \(D\) never vanishes, the transfer matrix
	\(\Xi_G(s)\) is  not strictly proper and therefore does not belong to
	\(\mathcal{RH}_2\). In the present work, however, the reduced-order model is always
	constructed with the same feed-through matrix, i.e.,
	\[
	D_r=D,
	\]
	so that the error transfer matrix
	\[
	\Xi_G(s)-\Xi_{G_r}(s)
	=
	C(sI-A)^{-1}B - C_r(sI-A_r)^{-1}B_r
	\]
	is strictly proper,  where \(\Xi_{G_r}(s)\) is the transfer matrix of the reduced-order model to be given in Section \ref{sec-H2-model-red}. Consequently, if both \(A\) and \(A_r\) are Hurwitz, then
	\(\Xi_G-\Xi_{G_r}\in\mathcal{RH}_2\), and the approximation error is measured by the square of the \(\mathcal{H}_2\) norm
\begin{multline}
	\|\Xi_G-\Xi_{G_r}\|_{\mathcal H_2}^2
	:=
	\frac{1}{2\pi}
	\int_{-\infty}^{\infty}
	\mathrm{trace}\!\left(
	\bigl(\Xi_G(i\omega)-\Xi_{G_r}(i\omega)\bigr)^*
	\right. \\
	\left.
	\bigl(\Xi_G(i\omega)-\Xi_{G_r}(i\omega)\bigr)
	\right)\,d\omega.
	\label{eq:H2_error_norm}
\end{multline}

	\subsection{Canonical structure and physical realizability}

	As the system \eqref{eQuantumSystem}  is a quantum linear system, it describes the Heisenberg equation of system observables \(x\) driven by bosonic field. Thus, the systems matrices \((A,B,C,D)\) cannot be arbitrary; instead they must satisfy the so-called physical realizability condition as coined in \cite{JamesNurdinPetersen2008}.

	For any integer \(k\ge 1\), define the canonical symplectic matrix
	\[
	J_k := I_k \otimes
	\begin{bmatrix}
		0 & 1\\
		-1 & 0
	\end{bmatrix}
	= I_k\otimes J_2.
	\]

	\begin{definition}[Physical realizability]
		The system \eqref{eQuantumSystem} is said to be physically realizable if it
		describes the Heisenberg evolution of a set of open quantum harmonic oscillators driven by
		bosonic fields; equivalently, if there exists a real symmetric matrix
		\(R=R^\top\in\mathbb{R}^{2n\times 2n}\) such that the dynamics are generated by
		the quadratic Hamiltonian
		\[
		H=\tfrac12 x^\top R x,
		\]
		with linear coupling to the bosonic fields, and the canonical commutation relations
		\begin{equation}\label{eq:CCR}
			[x(t),x(t)^\top]= i J_n
		\end{equation}
		are preserved for all time, and the system variables and output field quadratures satisfy the quantum non-demolition condition (causality)
		\begin{equation} \label{eq:nondemolition}
			[x_i(t), y_j(r)]=0,  \ 0\leq r\leq t,  \ \forall i=1,\ldots, 2n,j=1,\ldots 2m.
		\end{equation}
	\end{definition}

	Mathematically,  the
	physical realizability of the system \eqref{eQuantumSystem}  is equivalent to the algebraic
	identities
	\begin{equation}
		\begin{aligned}
			A J_n + J_n A^\top + B J_m B^\top &= 0,\\
			J_n C^\top + B J_m D^\top &= 0,\\
			D J_m D^\top &= J_m .
		\end{aligned}
		\label{eq:PR_full}
	\end{equation}
	The first identity expresses preservation of the canonical commutation
	relations  \eqref{eq:CCR}, the second couples internal dynamics and input-output structure and guarantees the quantum non-demolition condition  \eqref{eq:nondemolition}, and
	the third requires the direct feed-through to preserve field commutation
	relations; see \cite{VPB94,BvHJ07,JamesNurdinPetersen2008,Petersen2016,ZhangDong2022}.

	Several equivalent PR parametrizations for linear quantum systems appear in the literature, depending on
	the choice of coordinates and conventions; see, e.g.,
	\cite{JamesNurdinPetersen2008,Petersen2016,ZhangDong2022}. In the present
	quadrature convention, the first PR identity in \eqref{eq:PR_full} is
	equivalent to the existence of a real symmetric matrix
	$R=R^\top\in\mathbb{R}^{2n\times 2n}$ and a matrix \(B \in \mathbb{R}^{2n\times 2m}\) such that
	\begin{equation}
		A = J_n R + \tfrac12 B J_m B^\top J_n
		= J_n R - \tfrac12 B J_m B^\top J_n^\top .
		\label{eq:PR_param_standard}
	\end{equation}
	Indeed, setting $M:=B J_m B^\top$, we have $M^\top=-M$ because
	$J_m^\top=-J_m$. Hence,
	\[
	A = J_n R + \tfrac12 M J_n
	\]
	implies
	\[
	A J_n = J_n R J_n - \tfrac12 M,
	\qquad
	J_n A^\top = - J_n R J_n - \tfrac12 M,
	\]
	and therefore
	\[
	A J_n + J_n A^\top + B J_m B^\top
	=
	A J_n + J_n A^\top + M
	= 0.
	\]

	When $D=I_{2m}$, the second identity in \eqref{eq:PR_full} reduces to
	\[
	J_n C^\top + B J_m = 0,
	\]
	which is equivalent to
	\[
	C = J_m B^\top J_n.
	\]

	Thus, in the present convention, a convenient PR-by-construction quadrature
	template is given by
	\begin{equation}
		\begin{aligned}
			A &= J_n R + \tfrac12 B J_m B^\top J_n,\\
			C &= J_m B^\top J_n,\\
			D &= I_{2m},
		\end{aligned}
		\label{eq:PR_template}
	\end{equation}
	with $R=R^\top$.
	Indeed, \eqref{eq:PR_template} implies the physical realizability identities
	\eqref{eq:PR_full}.

	For later reference, we introduce the full-order PR residuals
	\begin{align}
		\mathcal R_1(A,B)
		:=&\; A J_n + J_n A^\top + B J_m B^\top, \notag \\
		\mathcal R_2(B,C,D)
		:=&\;  J_n C^\top + B J_m D^\top, \label{eq:PR_residuals_full} \\
		\mathcal R_3(D)
		:=&\;  D J_m D^\top - J_m . \notag
	\end{align}
	Thus a quadruple $(A,B,C,D)$ is physically realizable if and only if
	\[
	\mathcal R_1(A,B)=0,
	\qquad
	\mathcal R_2(B,C,D)=0,
	\qquad
	\mathcal R_3(D)=0.
	\]
	In particular, under the template \eqref{eq:PR_template}, all three residuals
	vanish identically.

	\section{The ${\mathcal H_2}$ model reduction problem}
	\label{sec-H2-model-red}

	Our goal is to approximate the full-order system \eqref{eQuantumSystem} by a reduced-order
	physically realizable system
	\begin{equation}
		\dot x_r = A_r x_r + B_r w,
		\qquad
		y_r = C_r x_r + D_r w,
		\label{eq:reduced_system}
	\end{equation}
	of order \(2r\ll 2n\), where
	\[
	A_r\in\mathbb{R}^{2r\times2r},
	B_r\in\mathbb{R}^{2r\times2m},
	C_r\in\mathbb{R}^{2m\times2r},
	D_r\in\mathbb{R}^{2m\times2m}.
	\]
	The corresponding reduced-order transfer matrix is given by
	\[
	\Xi_{G_r}(s)=C_r(sI-A_r)^{-1}B_r+D_r .
	\]

	The objective is to minimize the \(\mathcal H_2\) error between the full- and
	reduced-order transfer matrices while preserving physical realizability and
	stability. Since the feed-through term is retained unchanged, the reduction
	problem is formulated as
	\begin{equation}
		\min_{\substack{(A_r,B_r,C_r,D_r) \\
				\text{PR, } A_r \text{ Hurwitz},\ D_r=D}}
		\;
		\|\Xi_G - \Xi_{G_r}\|_{\mathcal H_2}.
		\label{eq:H2_min_problem}
	\end{equation}
	This problem is nonconvex \cite{WuXueZhangPetersen2025}, and in practice one seeks locally optimal
	solutions satisfying first-order optimality conditions of $\mathcal{H}_2$ optimal model reduction, which is recalled below.

	Let $H(s),\widehat{H}(s)\in \mathcal{RH}_2$ be such that $\widehat{H}(s)$ admits the
	pole-residue representation
	\[
	\widehat{H}(s) = \sum_{i=1}^r \frac{c_i b_i^*}{s-\lambda_i},
	\]
	where the poles $\lambda_i$ are pairwise distinct.
	If $\widehat{H}(s)$ is an $\mathcal{H}_2$-optimal approximation of $H(s)$, then for $i=1,\ldots,r$, the interpolation conditions
	\begin{align}
		H(-\bar{\lambda}_i)b_i &= \widehat{H}(-\bar{\lambda}_i)b_i, \notag \\
		c_i^*H(-\bar{\lambda}_i) &= c_i^*\widehat{H}(-\bar{\lambda}_i), \label{eq:necessary}\\
		c_i^*H'(-\bar{\lambda}_i)b_i &= c_i^*\widehat{H}'(-\bar{\lambda}_i)b_i \notag
	\end{align}
	are satisfied; see, e.g., \cite{GugercinAntoulasBeattie2008}.

	For linear quantum systems, a pole-zero duality exists. Specifically, by
	\cite[{Proposition 3.2}]{DongZhangLeePetersen2026}, $\lambda_i$ is a pole of
	$\widehat{H}(s)$ if and only if $-\bar{\lambda}_i$ is a transmission zero of
	$\widehat{H}(s)$. Consequently, the interpolation points appearing in
	\eqref{eq:necessary} may be interpreted as transmission zeros of the reduced-order
	model.  This observation provides additional insight into the pole-mirroring strategy
	employed in Q-IRKA, although the present method is constructed as a
	structure-preserving interpolatory iteration rather than as a direct solver of
	the constrained $\mathcal{H}_2$ optimality system.

	\subsection{Structure-preserving projection}

	In order to preserve the PR structure, we restrict attention to reduced-order models
	obtained by projection onto a symplectic trial subspace.
	Let \(V\in\mathbb{R}^{2n\times 2r}\) be a full-column-rank matrix satisfying
	\begin{equation}
		V^\top J_n V = J_r .
		\label{eq:symp_trial}
	\end{equation}
	The columns of \(V\) span a \(2r\)-dimensional symplectic subspace
	\(\mathcal V := \operatorname{Range}(V)\subset\mathbb{R}^{2n}\).

	The   vector of the reduced -order system variables \(x_r(t)\) represents the coordinate
	vector associated with the symplectic trial subspace \(\mathcal V\).
	The vector of  the full-order system variable is approximated by the ansatz
	\[
	x(t) \approx V x_r(t),
	\]
	so that model reduction is interpreted as the construction of an \emph{effective}
	low-dimensional symplectic coordinate representation rather than as an exact
	operator identity between canonical variables because in the quantum regime the latter is impossible when $r<n$.

	The reduced dynamics are then defined by
	Petrov-Galerkin projection with test
	basis
	\begin{equation}
		U := J_r^{-1} V^\top J_n ,
		\label{eq:test_basis}
	\end{equation}
	which satisfies
	\begin{equation}
		U V = I_{2r}.
		\label{eq:left_inverse}
	\end{equation}
	Moreover, the above defining relations imply the identity
	\begin{equation}
		V J_r = J_n^\top U^\top .
		\label{eq:useful_identity}
	\end{equation}
	This relation will be used repeatedly in verifying preservation of
	physical realizability of the reduced-order system under projection. This symplectic projection has also been used in \cite{TechakesariNurdin2017} in the study of model reduction of linear quantum systems.

	The reduced matrices are therefore given by
	\begin{equation}
		A_r = U A V,
	~
		B_r = U B,
~
		C_r = C V,
~
		D_r = D.
		\label{eq:proj_red}
	\end{equation}

	\begin{lemma}
		If $(A,B,C,D)$ satisfies \eqref{eq:PR_full} and
		$V^\top J_n V = J_r$ with
		$U = J_r^{-1} V^\top J_n$,
		then the reduced-order model \eqref{eq:proj_red} satisfies
		\begin{equation}
			\begin{aligned}
				A_r J_r + J_r A_r^\top + B_r J_m B_r^\top & = 0, \\
				J_r C_r^\top + B_r J_m D_r^\top & = 0, \\
				D_r J_m D_r^\top &= J_m .
			\end{aligned}
			\label{eq:PR_reduced}
		\end{equation}

	\end{lemma}

	For later use in the numerical sections, we define the reduced-order PR residuals
	as follows
	\begin{align}
		\mathcal R_{1,r}(A_r,B_r)
		& := A_r J_r + J_r A_r^\top + B_r J_m B_r^\top, \notag \\
		\mathcal R_{2,r}(B_r,C_r,D_r)
		& := J_r C_r^\top + B_r J_m D_r^\top, \label{eq:PR_residuals_reduced} \\
		\mathcal R_{3,r}(D_r)
		& := D_r J_m D_r^\top - J_m . \notag
	\end{align}
	Hence, the reduced-order model is physically realizable if and only if
	\[
	\mathcal R_{1,r}(A_r,B_r)=0, ~
	\mathcal R_{2,r}(B_r,C_r,D_r)=0,~
	\mathcal R_{3,r}(D_r)=0.
	\]
	Under symplectic projection, these residuals vanish exactly in exact arithmetic
	and remain at the level of floating-point roundoff in computation. Thus physical realizability is preserved automatically by symplectic projection,
	and in the numerical experiments this property is monitored through the residuals
	\eqref{eq:PR_residuals_reduced}.

	In this framework, the central task becomes the construction of
	$V\in\mathbb{R}^{2n\times 2r}$
	with $V^\top J_n V = J_r$ spanning an effective approximation space.
	A scalable approach is to build a tangential rational Krylov candidate space
	associated with a prescribed set of shifts $\{\sigma_i\}_{i=1}^r$ and a prescribed
	set of tangential directions $\{t_{i,\ell}\in\mathbb{R}^{2m}\}$,
	$\ell=1,\dots,L$, for each shift $\sigma_i$. Specifically, we will form the candidate pool
	\begin{equation}
		\begin{aligned}
			W_{\mathrm{cand}}
			=
			\bigl[
			&(A-\sigma_1 I)^{-1} B t_{1,1},\ \dots,\ (A-\sigma_1 I)^{-1} B t_{1,L},\\
			&\dots,\\
			&(A-\sigma_r I)^{-1} B t_{r,1},\ \dots,\ (A-\sigma_r I)^{-1} B t_{r,L}
			\bigr],
		\end{aligned}
		\label{eq:raw_basis}
	\end{equation}
	where the shifts \(\sigma_i\) are chosen outside \(\mathrm{spec}(A)\) so that
	the shifted linear systems are well defined. This condition is always satisfied if both the original system and the reduced-order system are Hurwitz stable. In the Q-IRKA iteration below,
	the shifts are updated as mirror images of selected  poles of  the latest reduced-order system.

	The vectors $t_{i,\ell}$ select input directions (or, more generally,
	tangential interpolation directions) in the $2m$-dimensional input space.
	In the numerical experiments, a deterministic pool is employed by choosing
	$t_{i,\ell}=e_{\ell}$ for $\ell=1,\dots,L$, where $\{e_{\ell}\}$ denotes the canonical
	basis of $\mathbb{R}^{2m}$. The enrichment parameter
	$L=L_{\mathrm{per\ shift}}$ controls the number of directions used per shift. In principle, larger values of $L_{\mathrm{per\ shift}}$ provide a more
	redundant candidate pool and may improve robustness of the extraction step.
	However, our numerical experience indicates that a practical choice
	$L_{\mathrm{per\ shift}}=r$ already yields the same reduction accuracy as
	larger values such as $L_{\mathrm{per\ shift}}=2r$, while substantially
	reducing the computational cost.

	From the candidate pool \(W_{\mathrm{cand}}\), the objective is to construct a
	\(2r\)-column matrix \(W\in\mathbb{R}^{2n\times2r}\) such that
	\(W^\top J_n W\) is nonsingular; equivalently, \(W\) is of full column rank. Conceptually, this may be viewed as a
	selection problem on the candidate pool. In the actual implementation of the Q-IRKA,
	however, a more efficient and robust symplectic Gram-Schmidt-type construction is used:
	after \(J_n\)-orthogonalization of a candidate vector, it is paired with its
	symplectic conjugate \(J_n^\top v\). A final symplectic normalization then
	constructs \(V=W T\) satisfying \(V^\top J_n V = J_r\). For this purpose, define
	\begin{equation}
		S := W^\top J_n W.
		\label{eq:S_def}
	\end{equation}
	Assume that $W$ has full column rank. Then $S=W^\top J_n W$
	is nonsingular. Symplectic normalization then consists of finding
	$T\in\mathbb{R}^{2r\times 2r}$ satisfying
	\begin{equation}
		T^\top S T = J_r
		\label{eq:T_def}
	\end{equation}
	and defining $V=W T$, so that $V^\top J_n V = J_r$.
	Hence, symplectic normalization \eqref{eq:T_def} is the change of coordinates $V=WT$ that
	puts the restricted pairing $S=W^\top J_n W$ into canonical form, $V^\top J_n V = J_r$.

	Once the projection structure is fixed, the \(\mathcal H_2\) model reduction problem
	reduces to selecting suitable interpolation points and directions. This approach leads to
	a symplectic analogue of IRKA. The dominant computational cost lies in solving
	shifted linear systems to determine \((A-\sigma_i I)^{-1}B t_{i,\ell}\), while the
	remaining operations involve only dense matrices of dimension \(2r\) and are
	negligible when \(r\ll n\).

	Since the reduced-order system dimension is \(2r\), the shift update rule is
	formulated so that the interpolation set reflects the spectral symmetry of
	\(A_r\). This yields a PR-preserving interpolatory fixed-point iteration
	inspired by classical IRKA and consistent with pole-mirroring ideas used in
	\(\mathcal H_2\) model reduction; see
	\cite{ABG10,GugercinAntoulasBeattie2008,MBDG25}.

	\begin{remark}
		{\rm
			A related PR-preserving tangential interpolatory framework for linear quantum
			systems was developed in \cite{TechakesariNurdin2017}. There, the reduced-order model
			is obtained from a structure-preserving projection that enforces interpolation
			at selected points and along selected tangential directions. Those
			interpolation data are specified as design choices by minimizing some \(\mathcal{H}_2\) norm, and moreover,  the construction is
			not organized as a recursive reduced-pole update.  In contrast, in our quantum analogue
			of the classical IRKA,  the interpolation points are not fixed once and for all;
			instead, they are updated \emph{recursively} by a pole-mirroring rule derived from the
			current reduced-order model. After each update, a new reduced-order system is computed by an updated 
			symplectic projection, so that physical realizability is maintained throughout
			the entire iteration.
		}
	\end{remark}

	\section{Quantum IRKA (Q-IRKA)}
	\label{sec:symp_qirka}

	Since physical realizability is enforced through symplectic projection, the
	\(\mathcal H_2\) model reduction problem reduces to constructing effective symplectic
	approximation spaces.
	Two main strategies can be considered: direct optimization over the symplectic
	Stiefel manifold, in the spirit of structure-preserving and manifold-based
	reduction approaches \cite{YL99,AbsilMahonySepulchre2008,PengMohseni2016}, or
	shift-based interpolatory methods analogous to IRKA.
	In this work, we adopt the latter approach, which iteratively
	updates interpolation points and recomputes the reduced-order model by symplectic
	projection.

	In the following, we first outline the three main steps of the Q-IRKA
	iteration, then present the algorithm, and finally discuss the implementation
	details used in the numerical experiments.

	\emph{Main step 1: Forming the current reduced-order model.}  At iteration \(k\), given interpolation points
	\(\{\sigma_i^{(k)}\}_{i=1}^r\), we construct a tangential rational Krylov candidate
	pool by solving shifted linear systems of the form
	\[
	(A-\sigma_i^{(k)}I) z = B \, t_{i,\ell},
	\]
	for the prescribed shifts \(\sigma_i^{(k)}\) and tangential directions
	\(t_{i,\ell}\). From this candidate pool a symplectic basis
	\(V^{(k)}\in\mathbb{R}^{2n\times2r}\) is extracted and normalized so that
	\[
	(V^{(k)})^\top J_n V^{(k)} = J_r .
	\]
	The reduced-order model is then assembled by symplectic Petrov-Galerkin projection
	as follows:
	\[
	A_r^{(k)} = U^{(k)} A V^{(k)},
~
	B_r^{(k)} = U^{(k)} B,
~
	C_r^{(k)} = C V^{(k)},
	\]
	with
	\[
	U^{(k)} = J_r^{-1} (V^{(k)})^\top J_n .
	\]

	\emph{Main step 2: updating the shifts.}
	The reduced matrix \(A_r^{(k)}\in\mathbb{R}^{2r\times2r}\) has \(2r\) poles
	(counting multiplicity), which appear in real or complex-conjugate pairs.
	The shift set is updated using representatives of these pairs.
	Specifically, we select \(r\) poles \(\lambda_i^{(k)}\) such that
	\(\Im(\lambda_i^{(k)})\ge0\) and define
	\[
	\sigma_i^{(k+1)} = -\overline{\lambda_i^{(k)}}, \qquad i=1,\dots,r.
	\]
	This produces a fixed-size ordered shift vector at each iteration and preserves
	the conjugation symmetry naturally associated with the real reduced-order model.

	\emph{Main step 3: updating  tangential directions.}
	At each iteration \(k\), the candidate pool is formed from the shifts
	\(\sigma_i^{(k)}\) and tangential directions
	\(t_{i,\ell}\in\mathbb{R}^{2m}\), \(\ell=1,\dots,L\).
	In the numerical experiments, we use a deterministic canonical pool
	\begin{equation}
		t_{i,\ell}=e_{\nu(\ell)},~ \ell=1,\dots,L,
		~
		\nu(\ell)=1+\operatorname{mod}(\ell-1,2m),
		\label{eq:tangential_choice}
	\end{equation}
	where \(\{e_j\}_{j=1}^{2m}\) denotes the canonical basis of
	\(\mathbb{R}^{2m}\). Thus, choosing \(L>2m\), the canonical directions are reused cyclically.
	This choice ensures that all input channels are sampled uniformly while
	allowing the pool size \(L\) to scale independently of the input dimension.
	In particular, when \(m\ll r\), this provides a sufficiently rich candidate
	set for the symplectic extraction step without introducing additional
	parameters or randomization.

	The complete structure-preserving iteration is summarized in the following
	symplectic Quantum IRKA algorithm.

\begin{algorithm}[t]
	\caption{Symplectic Quantum IRKA (Q-IRKA)}
	\label{alg:symp_qirka}
	\begin{algorithmic}[1]
		\Require Full-order PR model $(A,B,C,D)$ satisfying \eqref{eq:PR_full};
		structure matrices $J_n,J_m$; reduced dimension $2r$; initial shifts
		$\{\sigma_i^{(0)}\}_{i=1}^r$ with $\sigma_i^{(0)}\notin\mathrm{spec}(A)$;
		enrichment parameter $L_{\mathrm{per\ shift}}$; tolerance $\varepsilon$
		\Ensure Reduced PR model $(A_r,B_r,C_r,D_r)$ \break of order $2r$

		\State $k \gets 0$
		\Repeat
		\State \textbf{Enriched tangential Krylov pool.}
		\Statex \hspace{1.5em} Choose tangential directions \break
		$\{t_{i,\ell}\in\mathbb{R}^{2m}\}_{\ell=1}^{L_{\mathrm{per\ shift}}}$
		(e.g.\ $t_{i,\ell}=e_\ell$) for each $i=1,\dots,r$
		\Statex \hspace{1.5em} Form
		\[
		\begin{aligned}
			W_{\mathrm{cand}}^{(k)}
			=
			\bigl[
			&(A-\sigma_1^{(k)}I)^{-1}B t_{1,1},\dots,(A-\sigma_1^{(k)}I)^{-1}B t_{1,L},\\
			&\dots,\\
			&(A-\sigma_r^{(k)}I)^{-1}B t_{r,1},\dots,(A-\sigma_r^{(k)}I)^{-1}B t_{r,L}
			\bigr].
		\end{aligned}
		\]

		\State \textbf{Symplectic extraction.}
		\Statex \hspace{1.5em} Apply a symplectic Gram-Schmidt-type procedure:
		after $J_n$-orthogonalization of a candidate vector, form a pair
		$(v,\,J_n^\top v)$ and append it to the basis.
		Continue until $2r$ columns are obtained, yielding
		$W^{(k)}\in\mathbb{R}^{2n\times 2r}$.

		\State \textbf{Symplectic normalization.}
		\Statex \hspace{1.5em} Form $S^{(k)}=(W^{(k)})^\top J_n W^{(k)}$
		and compute $T^{(k)}$ such that
		$(T^{(k)})^\top S^{(k)} T^{(k)}=J_r$
		\Statex \hspace{1.5em} Set $V^{(k)} = W^{(k)} T^{(k)}$ so that \break
		$(V^{(k)})^\top J_n V^{(k)}=J_r$

		\State \textbf{Left inverse.}
		\Statex \hspace{1.5em} $U^{(k)} \gets J_r^{-1}(V^{(k)})^\top J_n$

		\State \textbf{Reduced model (PR by construction).}
		\Statex \hspace{1.5em}
		$A_r^{(k)} \gets U^{(k)} A V^{(k)}$,
		$B_r^{(k)} \gets U^{(k)} B$, \break
		$C_r^{(k)} \gets C V^{(k)}$,
		$D_r \gets D$

		\State \textbf{Shift update.}
		\Statex \hspace{1.5em} Compute the reduced poles \break
		$\{\lambda_i^{(k)}\}_{i=1}^{2r}\subset\mathrm{spec}(A_r^{(k)})$
		(ordered consistently across iterations)
		\Statex \hspace{1.5em} Set $\sigma_i^{(k+1)} \gets -\lambda_i^{(k)}$
		for $i=1,\dots,r$ \break (using conjugate pairing if complex)

		\State \textbf{Stopping test.}
		\Statex \hspace{1.5em} $\mathrm{relchg} \gets
		\dfrac{\|\sigma^{(k+1)}-\sigma^{(k)}\|_2}{\max\{1,\|\sigma^{(k)}\|_2\}}$
		\State $k \gets k+1$
		\Until{$\mathrm{relchg} < \varepsilon$}

		\State \Return $(A_r,B_r,C_r,D_r) \gets (A_r^{(k)},B_r^{(k)},C_r^{(k)},D_r)$
	\end{algorithmic}
\end{algorithm}

	Algorithm~\ref{alg:symp_qirka} summarizes the conceptual iteration.
	The following implementation details specify the concrete procedures used in
	our numerical experiments.

	\smallskip

	\noindent\textbf{(1) Symplectic extraction from an enriched pool.}
	Let \(W_{\mathrm{cand}}^{(k)}=[w_1,\dots,w_M]\in\mathbb{R}^{2n\times M}\).
	We construct \(W^{(k)}\in\mathbb{R}^{2n\times 2r}\) incrementally using a
	symplectic Gram-Schmidt-type procedure. Set \(W^{(k)}=\emptyset\). While
	\(\dim(W^{(k)})<2r\), proceed as follows:
	\begin{enumerate}
		\item Select the next candidate vector \(w_j\) in \(W_{\mathrm{cand}}^{(k)}\) and compute its
		\(J_n\)-orthogonal residual
		\[
		\widehat w
		=
		w_j-
		W^{(k)}\bigl((W^{(k)})^\top J_n W^{(k)}\bigr)^{-1}(W^{(k)})^\top J_n w_j.
		\]
		\item If \(\|\widehat w\|_2\le \tau_{\mathrm{sgs}}\) which is set to be $10^{-12}$ in all experiments, discard \(w_j\) and continue.
		\item Otherwise, set
		\[
		v=\frac{\widehat w}{\|\widehat w\|_2},
		\qquad
		u=J_n^\top v,
		\]
		and append the pair
		\[
		W^{(k)} \leftarrow [\,W^{(k)}\; v\; u\,].
		\]
		For improved numerical stability, one optional re-orthogonalization pass is
		applied after the pair $(v,J_n^\top v)$ has been appended to the basis.
		More precisely, the two newly added columns are projected once more against the
		previously accepted columns with respect to the $J_n$-pairing and are then
		renormalized. This helps reduce loss of $J_n$-orthogonality caused by
		floating-point roundoff when candidate vectors become nearly dependent.
		In the reported experiments, one such re-orthogonalization pass was used.
	\end{enumerate}
	This construction does not enforce symplecticity explicitly, but produces a
	matrix \(W^{(k)}\) for which the pairing matrix
	\((W^{(k)})^\top J_n W^{(k)}\) is nonsingular in practice. The exact
	symplectic constraint is then imposed in a separate normalization step described below.

	\smallskip

	\noindent\textbf{(2) Symplectic normalization.} After a symplectic extraction step has selected $2r$ columns and produced
	$W^{(k)}\in\mathbb{R}^{2n\times 2r}$, the basis is mapped to a $J_r$-orthonormal
	(symplectic) form. To this end, the pairing matrix
	\begin{equation}
		S^{(k)} := (W^{(k)})^{\top} J_n W^{(k)} \in \mathbb{R}^{2r\times 2r}
		\label{eq:pairing_matrix}
	\end{equation}
	is formed. Since \(J_n^\top=-J_n\), the matrix \(S^{(k)}\) is skew-symmetric.
	If \(\mathrm{rank}(S^{(k)})=2r\), a change of coordinates
	\(T^{(k)}\in\mathbb{R}^{2r\times2r}\) can be computed such that
	\begin{equation}
		(T^{(k)})^\top S^{(k)} T^{(k)} = J_r,
		\qquad
		V^{(k)} = W^{(k)}T^{(k)}.
		\label{eq:symplectic_normalization}
	\end{equation}

	In computations, $T^{(k)}$ in Eq. \eqref{eq:symplectic_normalization} is obtained via
	a real Schur decomposition of the skew-symmetric matrix $S^{(k)}$ as follows:
	\[
	S^{(k)} = Q^{(k)} R^{(k)} \bigl(Q^{(k)}\bigr)^{\top},
	\]
	where $Q^{(k)}$ is orthogonal and $R^{(k)}$ is block diagonal with $2\times2$ blocks
	of the form
	\[
	\begin{bmatrix} 0 & \alpha_j \\ -\alpha_j & 0 \end{bmatrix},
	\qquad \alpha_j>0,
	\qquad j=1,\dots,r.
	\]
	A block-diagonal scaling
	\[
	D^{(k)} := \operatorname{blkdiag}\bigl(\alpha_1^{-1/2} I_2,\dots,\alpha_r^{-1/2} I_2\bigr)
	\]
	is then applied and the normalization matrix is set to
	\begin{equation}
		T^{(k)} := Q^{(k)} D^{(k)}.
		\label{eq:T_from_schur_scaling}
	\end{equation}
	With Eq. \eqref{eq:T_from_schur_scaling}, the identity \eqref{eq:symplectic_normalization}
	is satisfied and the symplectic constraint $\bigl(V^{(k)}\bigr)^{\top} J_n V^{(k)}=J_r$
	is enforced in floating point arithmetic.

	\smallskip
	\noindent\textbf{(3) Complex shifts and real basis construction.}
	When a complex shift $\sigma=a+ib$ with $b\neq 0$ occurs, the corresponding
	complex Krylov vector
	\[
	z(\sigma,t) := (A-\sigma I)^{-1}B t \in \mathbb{C}^{2n}
	\]
	is not stored directly. Instead, for each tangential direction $t$, we insert
	the two real vectors
	\[
	\Re\bigl(z(\sigma,t)\bigr), \qquad \Im\bigl(z(\sigma,t)\bigr)
	\]
	into the candidate pool. \\
	Since the full-order matrices are real, these two vectors span the same real
	subspace associated with the conjugate pair $\sigma,\overline{\sigma}$.
	This ensures that $W_{\mathrm{cand}}^{(k)}$ (and hence $W^{(k)}$) remains
	real-valued even when complex interpolation points occur.

	\smallskip
	\noindent\textbf{(4) Pole-to-shift selection for a $2r$-dimensional reduced model.}
	At iteration $k$, the reduced matrix $A_r^{(k)}\in\mathbb{R}^{2r\times 2r}$ has
	$2r$ eigenvalues (counting multiplicity) that appear in complex-conjugate pairs.
	We form the next shift set $\{\sigma_i^{(k+1)}\}_{i=1}^r$ by selecting one
	representative from each pair, namely
	\[
	\sigma_i^{(k+1)} \;=\; -\lambda_i^{(k)},
	\qquad
	\text{where }\Im(\lambda_i^{(k)})\ge 0,
	\quad i=1,\dots,r,
	\]
	and the eigenvalues are sorted by increasing imaginary part and, within ties,
	by increasing real part. This convention yields a unique ordered vector
	$\sigma^{(k)}\in\mathbb{C}^r$ used in the stopping test.

	\smallskip
	\noindent\textbf{(5) Stopping criterion.}
	With the above ordering, the relative shift change is computed as
	\[
	\mathrm{relchg}^{(k)}
	=
	\frac{\|\sigma^{(k+1)}-\sigma^{(k)}\|_2}
	{\max\{1,\|\sigma^{(k)}\|_2\}},
	\]
	and the iteration is terminated once \(\mathrm{relchg}^{(k)}<\varepsilon\).

	For the numerical assessment of reduction quality, we use standard
	Gramian-based diagnostics. For the original full-order system \eqref{eQuantumSystem},
	since $A$ is Hurwitz, the controllability and observability Gramians are obtained as the unique solutions
	of the continuous-time Lyapunov equations
	\begin{align*}
		A W_c + W_c A^\top + B B^\top =&\; 0, \\
		A^\top W_o + W_o A + C^\top C &\;= 0.
	\end{align*}
	The Hankel singular values are then computed from the eigenvalues of
	$W_c^{1/2} W_o W_c^{1/2}$.
	These quantities are used to generate the Gramian spectra and the
	Hankel singular value plots reported in the numerical sections.
	In addition, the $\mathcal H_2$ norm of a strictly proper transfer matrix is
	computed as,
	\[
	\|\Xi_G(s)-D\|_{\mathcal H_2}^2 = \operatorname{trace}(C W_c C^\top).
	\]
	Assuming that the reduced matrix $A_r$ is Hurwitz, the approximation
	error $\|\Xi_G(s) - \Xi_{G_r}(s)\|_{\mathcal H_2}$ can be evaluated in a similar way
	by solving a Lyapunov equation for the associated error system $\Xi_G(s) - \Xi_{G_r}(s)$.
	As this procedure is standard, the details are omitted; see \cite[Eqs. (10)-(14)]{WuXueZhangPetersen2025} or \cite[Eqs. (1)-(7)]{YL99}.

	In the following two sections, we define deterministic physically motivated
	benchmarks for large-scale testing
	and discuss the observed numerical performance of our method. For later use in the numerical sections, we define the structural diagnostics
	\begin{equation}
		\begin{aligned}
			\Delta_{\mathrm{symp}} &:= \|V^\top J_n V - J_r\|_F,\\
			\Delta_{\mathrm{left}} &:= \|U V - I_{2r}\|_F,\\
			\Delta_{\mathrm{PR},j} &:= \|\mathcal R_{j,r}\|_F,
			\quad j=1,2,3.
		\end{aligned}
		\label{eq:structural_diagnostics}
	\end{equation}
	When the feed-through is fixed as $D_r=D=I$, one has
	$\mathcal R_{3,r}=0$ identically, so only
	$\Delta_{\mathrm{PR},1}$ and $\Delta_{\mathrm{PR},2}$ need to be reported.

	We end this section by comparing the proposed Q-IRKA method with the quasi-balanced truncation  model reduction method \cite{Nurdin14} and the non-convex optimization method \cite{WuXueZhangPetersen2025}, by means of a  small-scale example.

\begin{example}
	{\rm
		We  consider the independent-oscillator realization (a bus model) of quantum linear system as shown in \cite[Fig.~2]{GZ15}. In our experiment,  $n=1$ and $m=1$, i.e., this system has 10 system modes and is driven by a single bosonic field. For the model (30)-(37) in Theorem 5 of \cite{GZ15}, we chose  $\gamma =2.2$, $\omega_0=1.0$,  $\omega_j$, $(j=1,\ldots,9)$, are chosen sequentially  as $\{4.18, 3.28, 2.42, 2.28, 1.75, 1.61,1.55, 1.40, 1.20\}$ and $\kappa_j$, $(j=1,\ldots,9)$,  as $\{0.95, 0.78, 0.66, 0.58, 0.44, 0.31$, $0.22, 0.14, 0.08\}$.     We reduced this 20-dimensional quantum system to a 6-dimensional reduced-order model.  The $\mathcal{H}_2$ error is  1.86 obtained using the quasi-balanced truncation  model reduction method in \cite{Nurdin14}, and it is  1.25 by means of the method in \cite{WuXueZhangPetersen2025}. In contrast, the  $\mathcal{H}_2$ error  is  1.2130  by the proposed Q-IRKA method, which is the best among the three.
	}
\end{example}

	The full-order model in the above example is admittedly small. In the following two sections, we apply the Q-IRKA algorithm to two large-scale benchmarks.

	\section{Low-channel port-Hamiltonian}
	\label{sec:lowchannel_detailed}

	We consider a low-channel oscillator-chain benchmark with $m<n$, where
	multiple oscillators share common dissipative pathways. To preserve
	physical realizability, a PR-by-construction full-port dilation is used,
	following structure-preserving port-Hamiltonian principles
	\cite{PolyugavanDerSchaft2010}. External channels define the observable
	input-output behaviour, while additional site-wise channels ensure
	internal dissipation.

	Let \(R=R^\top\succeq 0\) denote the symmetric energy matrix in the quadratic
	Hamiltonian
	\[
	H(x)=\tfrac12 x^\top R x .
	\]
	Let
	\[
	\kappa_{\mathrm{ch}}\in\mathbb{R}_+^m,
	\qquad
	\kappa_{\mathrm{site}}\in\mathbb{R}_+^n,
	\]
	with positive entries. The vector \(\kappa_{\mathrm{ch}}\) contains the damping
	strengths of the observable external channels, whereas
	\(\kappa_{\mathrm{site}}\) contains the site-wise hidden damping strengths. \\
	Channel aggregation is described by an attachment matrix
	\(S\in\mathbb{R}^{n\times m}\). The entry \(S_{j\ell}\) specifies the attachment
	of the \(\ell\)-th external channel to the \(j\)-th oscillator. We define
	\begin{align*}
	B_{\mathrm{ch}}
	=&\;
	\bigl(S\operatorname{diag}(\sqrt{\kappa_{\mathrm{ch}}})\bigr)\otimes I_2
	\in\mathbb{R}^{2n\times 2m},
	\\
	B_{\mathrm{site}}
	=&\;
	\operatorname{diag}(\sqrt{\kappa_{\mathrm{site}}})\otimes I_2
	\in\mathbb{R}^{2n\times 2n}.
	\end{align*}
	Thus, the full-port coupling and structure matrices are given by
	\begin{align*}
	&B_{\mathrm{tot}} = [B_{\mathrm{ch}},\,B_{\mathrm{site}}],
~
	J_{\mathrm{tot}} = \operatorname{blkdiag}(J_m,J_n),
	\\
 & D_{\mathrm{tot}} = I_{2(m+n)}.
	\end{align*}

	The full-port system is given by
	\begin{equation}
		\begin{aligned}
			A &= J_n R + \tfrac12 B_{\mathrm{tot}} J_{\mathrm{tot}} B_{\mathrm{tot}}^\top J_n,\\
			C_{\mathrm{tot}} &= J_{\mathrm{tot}} B_{\mathrm{tot}}^\top J_n,
		\end{aligned}
		\label{eq:benchmark2_fullport_template}
	\end{equation}
	and satisfies the PR conditions up to machine precision.

	Two configurations are considered. In the homogeneous case,
	$\max \Re(\lambda(A)) = -10^{-1}$, while in the heterogeneous case
	$\max \Re(\lambda(A)) = -10^{-2}$, indicating weaker damping and
	higher dynamical complexity.

\begin{figure}[H]
		\centering
		\includegraphics[width=0.48\linewidth]{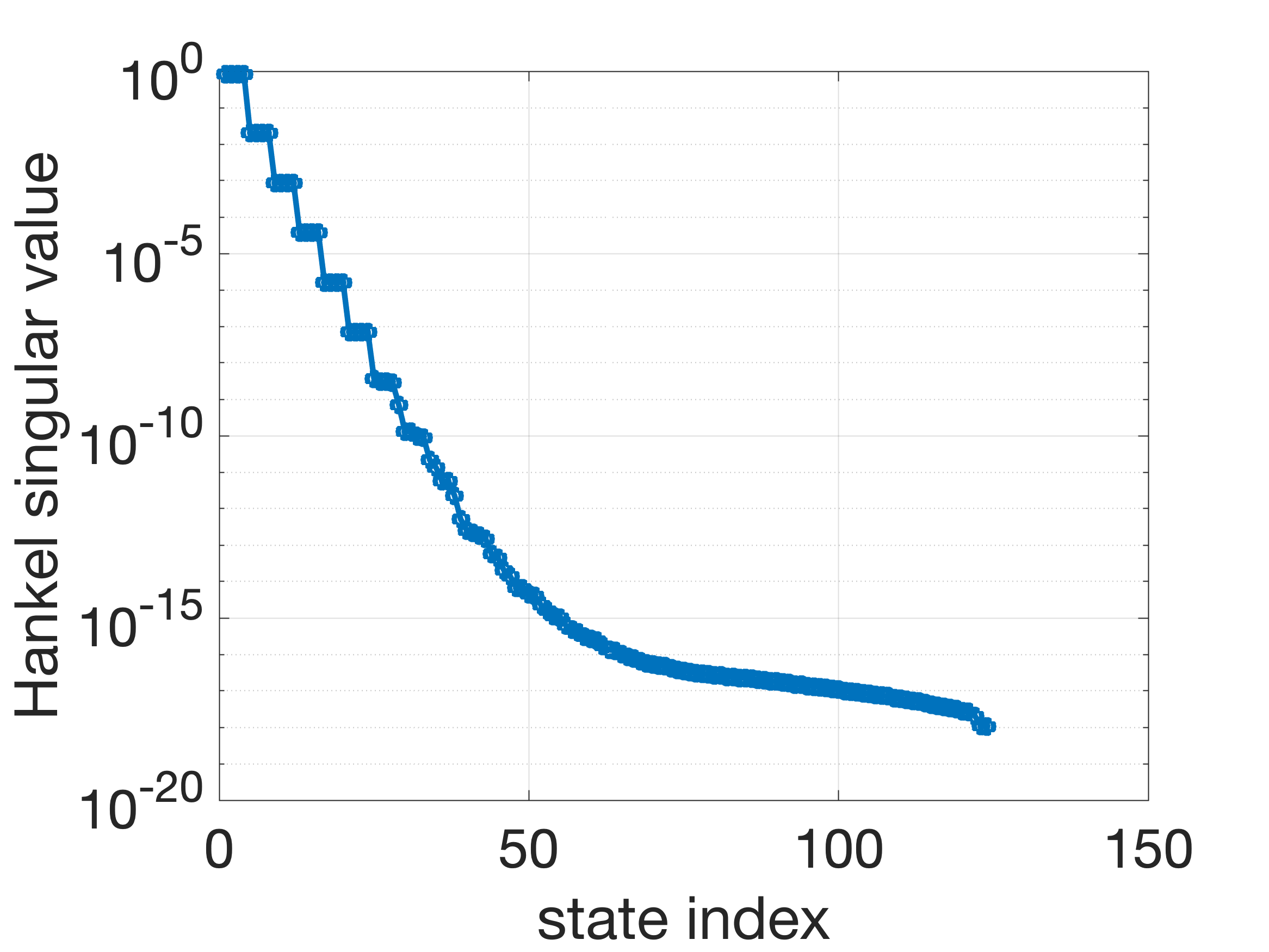}
		\includegraphics[width=0.48\linewidth]{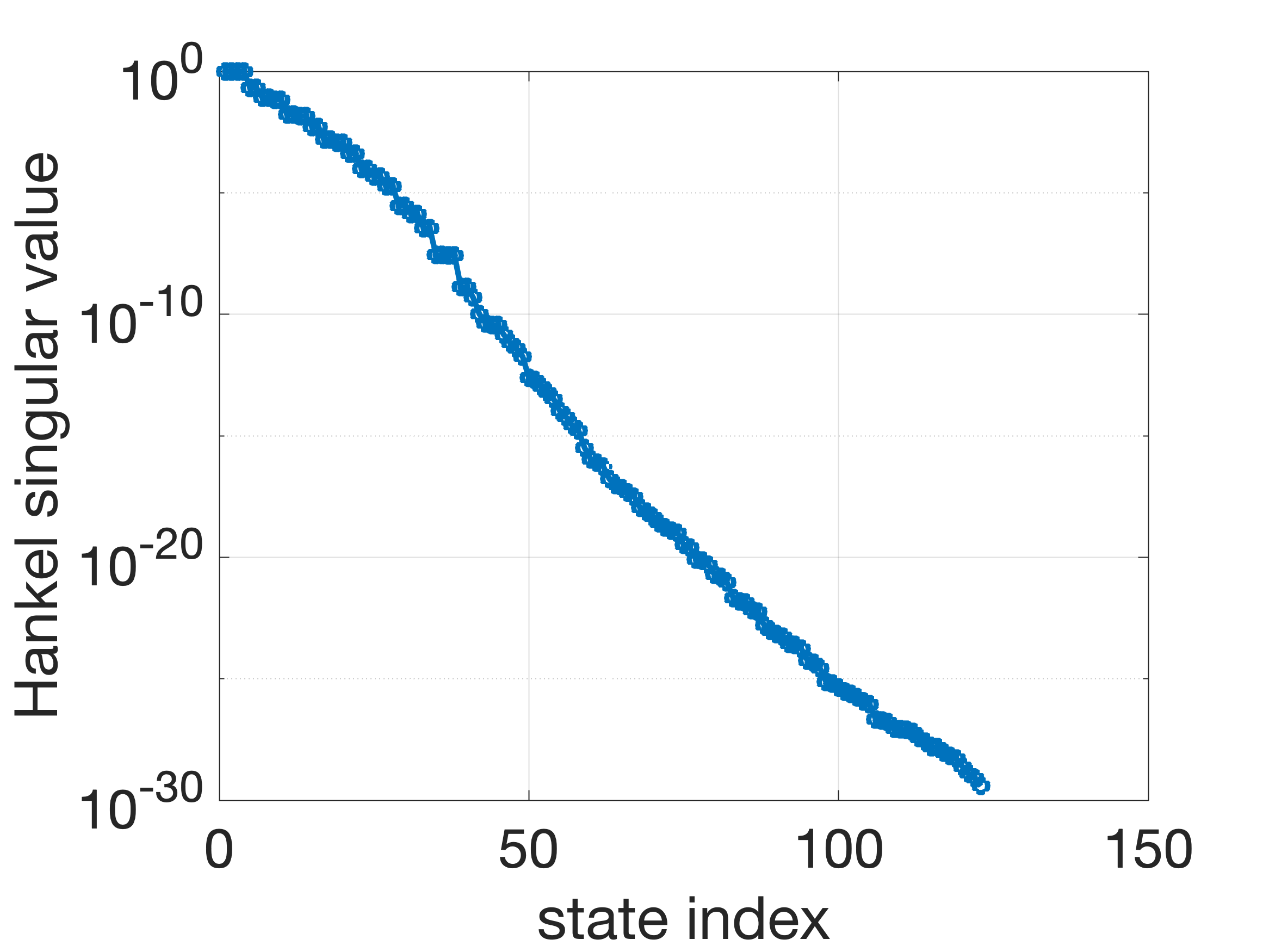}
		\caption{Hankel singular values; homogeneous (left) and heterogeneous (right).}
		\label{fig:hsv_hom}
	\end{figure}

	We analyze the case $(n,m,r)=(100,2,10)$.
	The relation between Hankel singular values and approximation error is classical;
	see, e.g., \cite{Antoulas2005}.
	The Hankel singular values decay smoothly in both configurations without
	visible clustering. In the homogeneous case,
	\[
	\sigma_{10} \approx \mathcal{O}(10^{-3}),
	\]
	while in the heterogeneous case
	\[
	\sigma_{10} \approx \mathcal{O}(10^{-2}),
	\]
	indicating slower decay and a higher effective dynamical rank; see
	Figure~\ref{fig:hsv_hom}.

\begin{figure}[H]
		\centering
		\includegraphics[width=0.48\linewidth]{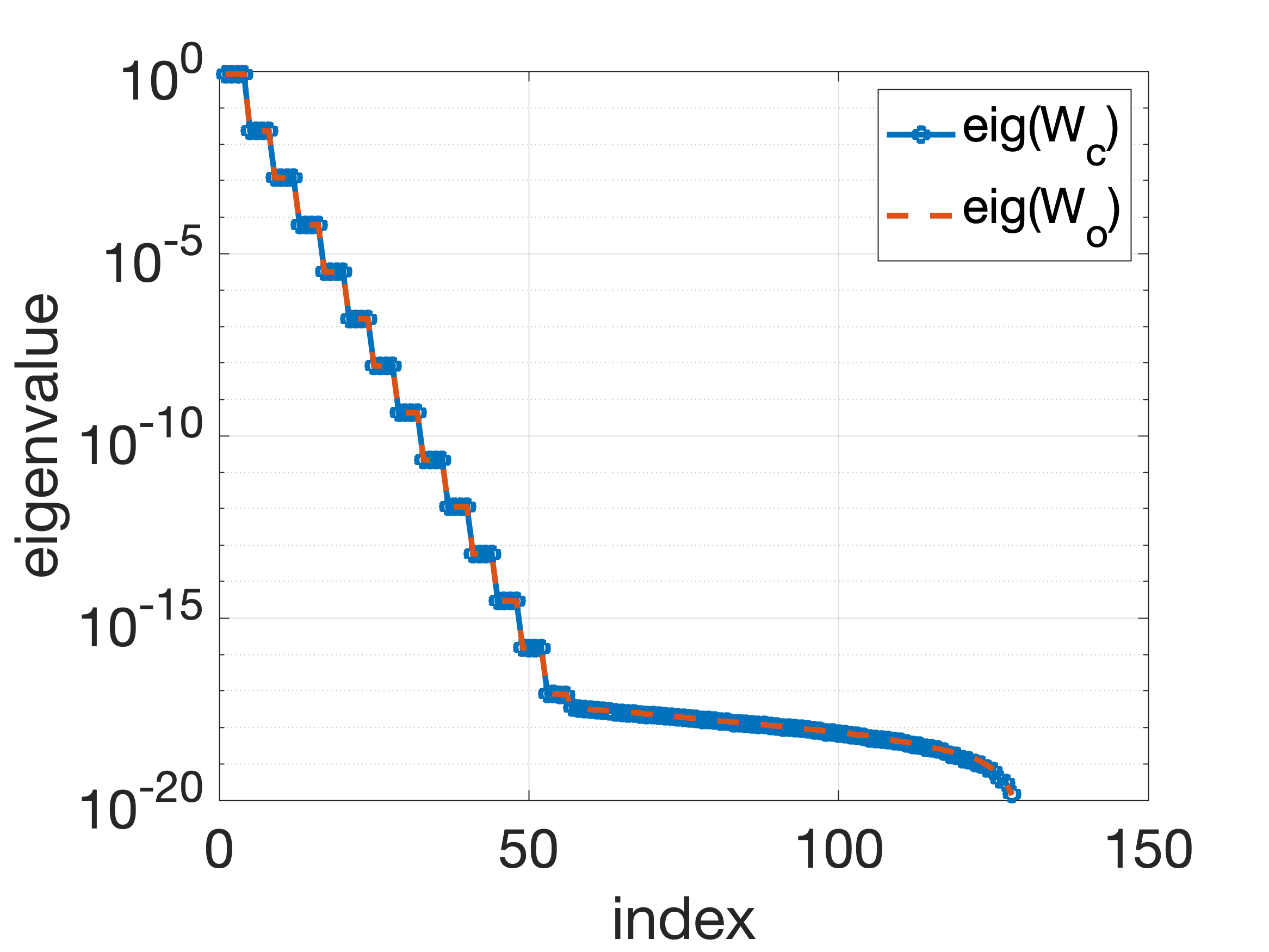}
		\includegraphics[width=0.48\linewidth]{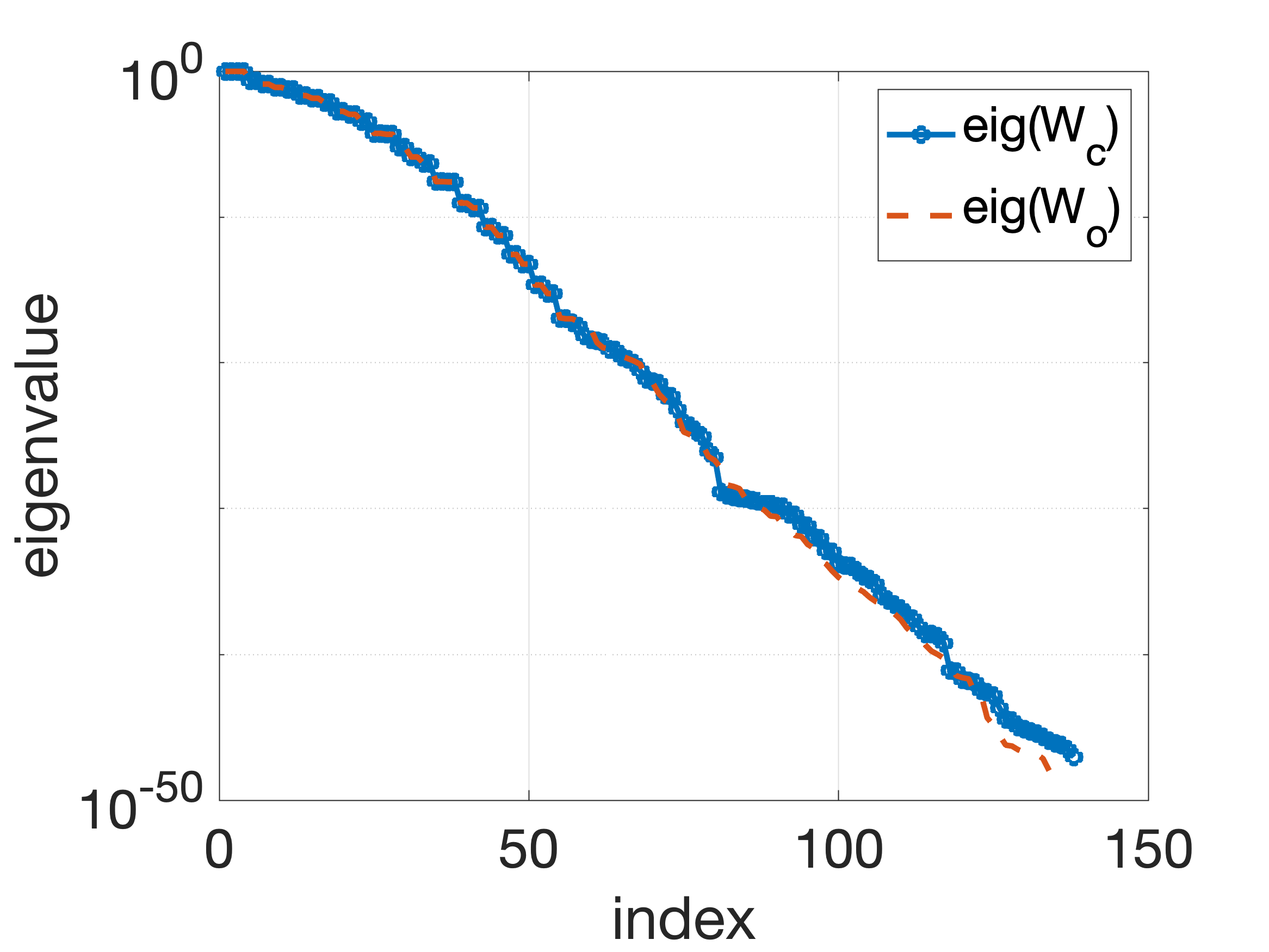}
		\caption{Gramian spectra; homogeneous (left) and heterogeneous (right).}
		\label{fig:gram_hom}
	\end{figure}

	The Gramian spectra satisfy $\operatorname{eig}(W_c)\approx\operatorname{eig}(W_o)$,
	reflecting the underlying PR symmetry, and exhibit smooth decay without
	pronounced gaps; see Figure~\ref{fig:gram_hom}.

\begin{figure}[H]
		\centering
		\includegraphics[width=0.48\linewidth]{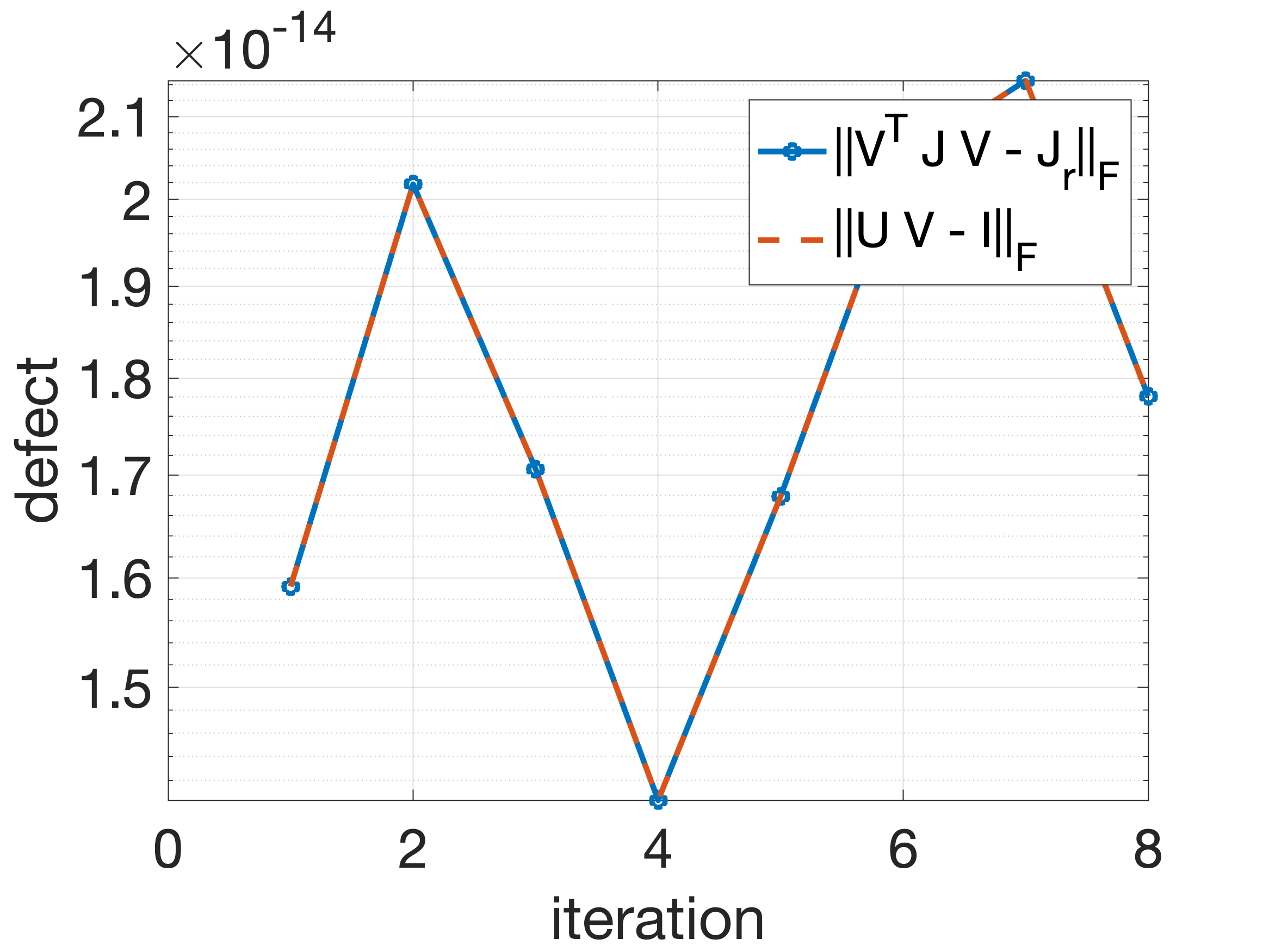}
		\includegraphics[width=0.48\linewidth]{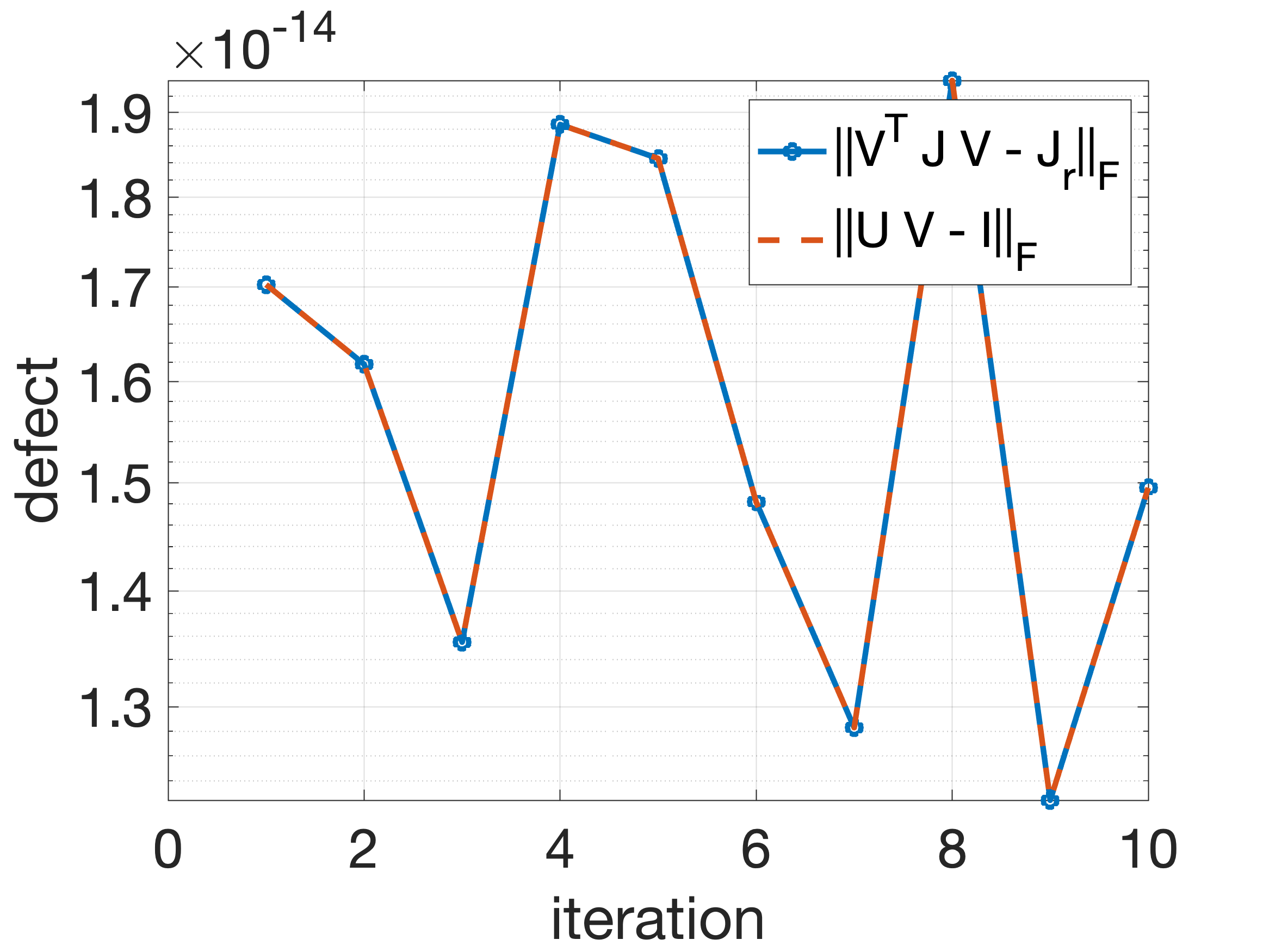}
		\caption{Symplectic diagnostics; homogeneous (left) and heterogeneous (right).}
		\label{fig:symp_hom}
	\end{figure}

	Q-IRKA is applied to the external-port system
	$(A,B_{\mathrm{ext}},C_{\mathrm{ext}},D_{\mathrm{ext}})$, while PR
	structure is monitored on the full-port system. The structural defects
	remain at machine precision,
	\[
	\Delta_{\mathrm{symp}} \approx 10^{-14},
	\qquad
	\Delta_{\mathrm{PR},j} \approx 10^{-16},
	\]
	throughout the iteration; see Figure~\ref{fig:symp_hom}.

\begin{figure}[H]
		\centering
		\includegraphics[width=0.48\linewidth]{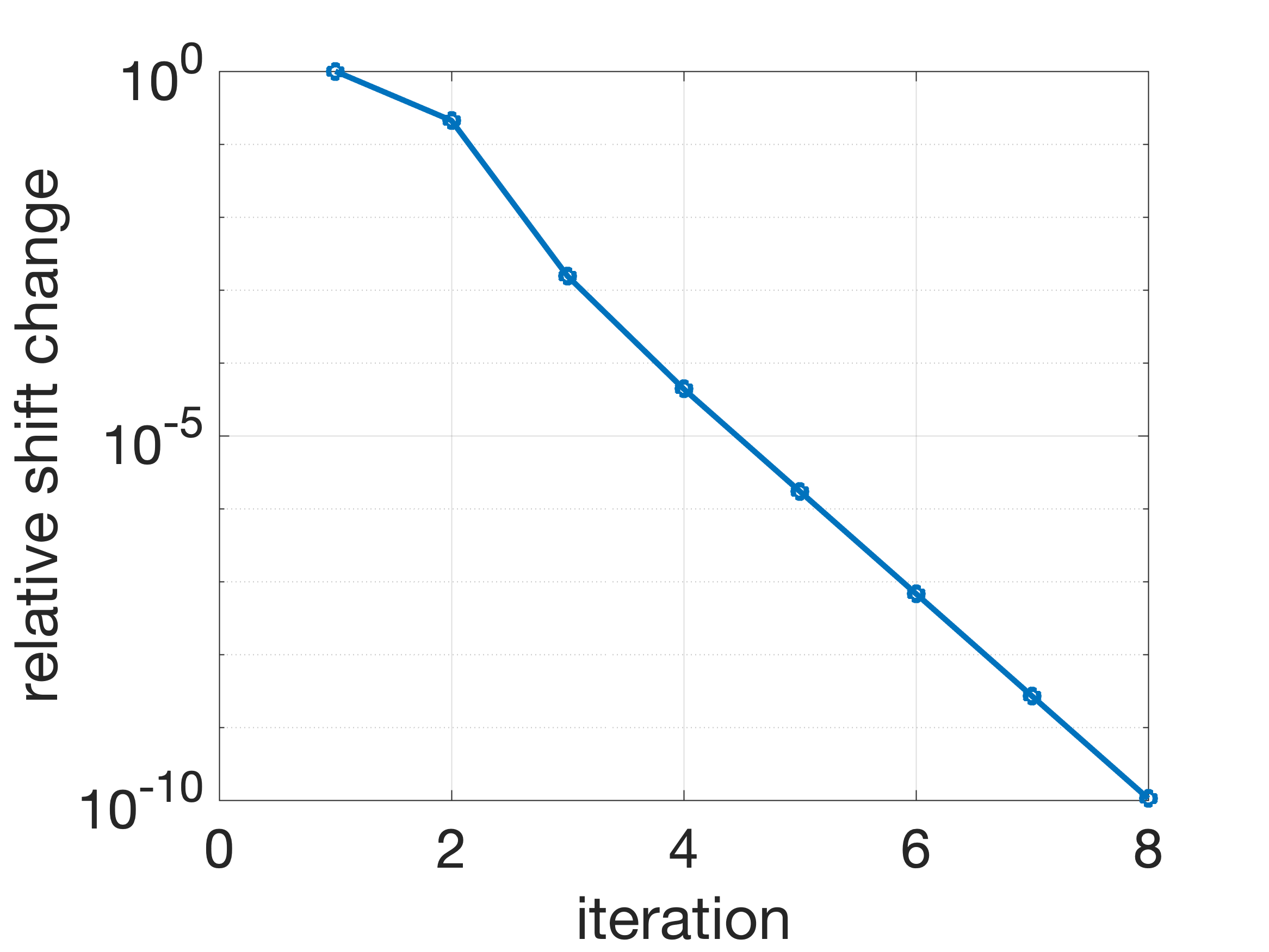}
		\includegraphics[width=0.48\linewidth]{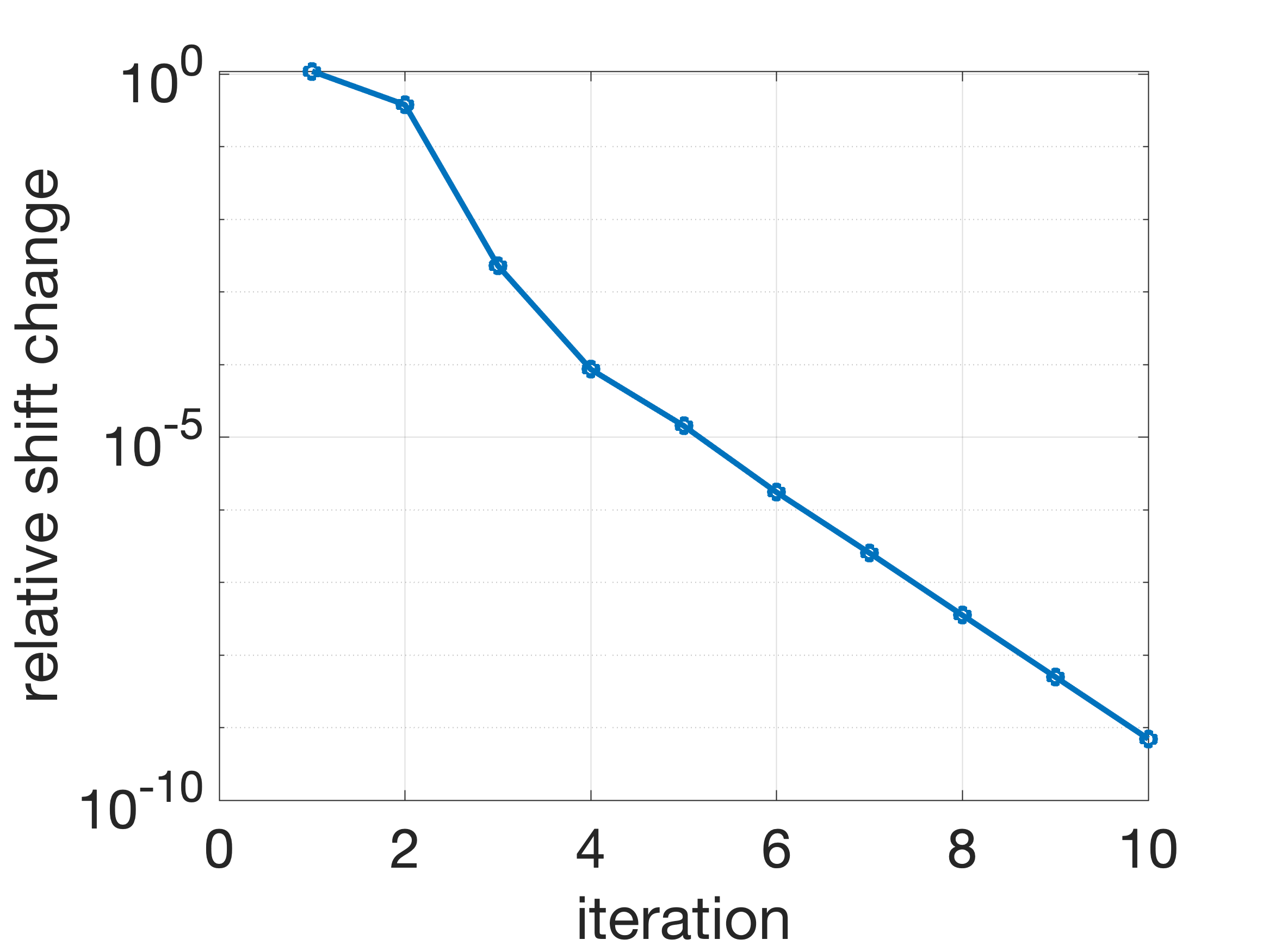}
		\caption{Shift convergence; homogeneous (left) and heterogeneous (right).}
		\label{fig:shift_hom}
	\end{figure}

	The shift iteration converges monotonically in both configurations,
	without oscillations, indicating well-separated reduced poles;
	see Figure~\ref{fig:shift_hom}.

	The achieved $\mathcal{H}_2$ errors reflect the spectral decay:
	the homogeneous case admits accurate low-order approximation,
	whereas the heterogeneous case is intrinsically harder due to slower
	Hankel decay.

	For complexity, we consider the set
	\[
	(n,m,r) \in \left\{
	\begin{array}{l}
		(50,2,10),\,
		(100,2,10),\,
		(200,2,10),\\
		(100,4,10),\,
		(200,4,10),\,
		(200,4,20)
	\end{array}
	\right\},
	\]
	see Table~\ref{tComplexityQIRKA01}.

	\begin{table*}[t]
		\centering

		\caption{Runtime and relative $\mathcal{H}_2$ error.}

		\begin{tabular}{c c c | c c | c c}
			\hline
			$n$ & $m$ & $r$
			& $T_{\text{hom}}$ & $E_{\text{hom}}$
			& $T_{\text{het}}$ & $E_{\text{het}}$ \\
			\hline
			50 & 2 & 10 & 0.235 & 2.7e-05 & 0.177 & 7.0e-04 \\
			100 & 2 & 10 & 0.678 & 2.7e-05 & 0.854 & 5.2e-03 \\
			200 & 2 & 10 & 2.210 & 2.7e-05 & 4.224 & 1.3e-02 \\
			100 & 4 & 10 & 0.506 & 2.7e-05 & 0.561 & 5.7e-03 \\
			200 & 4 & 10 & 1.684 & 2.7e-05 & 2.809 & 1.1e-02 \\
			200 & 4 & 20 & 61.794 & 3.9e-09 & 14.941 & 4.8e-03 \\
			\hline
		\end{tabular}
		\label{tComplexityQIRKA01}
	\end{table*}

	Table~\ref{tComplexityQIRKA01} is best read by varying one parameter at a time.
	For fixed \(m\) and \(r\), the runtime increases with \(n\); for fixed \(n\)
	and \(m\), it increases more markedly with \(r\). The heterogeneous cases also
	tend to exhibit larger errors and often higher computational cost.
	The dependence on \(m\) is mild, while the cost increases with \(r\) due to
	larger projection spaces and an increased number of interpolation conditions.
	The error behaviour is governed by Hankel decay, which is essentially independent
	of \(n\) in the homogeneous case, but increases with \(n\) in the heterogeneous
	case. Increasing \(r\) significantly improves accuracy, especially when
	spectral decay is fast.

	\section{Bosonic Kitaev-Majorana chain}
	\label{sec:bkc_benchmark}

	To assess Q-IRKA on a structured transport-dominated benchmark, we consider a
	family of quantum systems inspired by the bosonic
	Kitaev-Majorana chain realized experimentally in \cite{Slim2024}. Our purpose
	is not to reproduce the full optomechanical model, but to retain the features
	most relevant for model reduction: nearest-neighbour coupling,
	quadrature-resolved transport, boundary-driven input-output behaviour, and a
	low-channel architecture.

	We consider a chain of \(n\) bosonic modes in canonical quadrature coordinates
	\[
	x=(x_1,p_1,\dots,x_n,p_n)^\top ,
	\]
	with quadratic Hamiltonian
	\[
	H(x)=\tfrac12 x^\top R x,
	\]
	where \(R=R^\top\in\mathbb{R}^{2n\times 2n}\) is block tridiagonal,
	\[
	R =
	\begin{bmatrix}
		R_0^{(1)} & R_1^\top &        &        \\
		R_1       & R_0^{(2)} & R_1^\top &      \\
		& \ddots   & \ddots & \ddots \\
		&          & R_1    & R_0^{(n)}
	\end{bmatrix},
	\qquad
	R_0^{(j)}=\omega_j I_2,
	\]
	with nearest-neighbour coupling block
	\[
	R_1=
	\begin{bmatrix}
		0 & \beta\\
		\alpha & 0
	\end{bmatrix},
	\qquad
	\alpha = |\lambda|-|J|,
	\qquad
	\beta = |\lambda|+|J|.
	\]
	In the balanced regime \(|J|=|\lambda|\), one has \(\alpha=0\), giving the
	directional quadrature-coupling pattern associated with the bosonic
	Kitaev-Majorana chain in \cite{Slim2024}. In our experiments, we consider a
	nearly balanced regime, so that \(|\alpha|\) is small and the coupling remains
	anisotropic.

	To obtain a stable low-channel physically realizable benchmark, we use the same
	full-port dilation strategy as in Section~\ref{sec:lowchannel_detailed}. The
	full-port coupling matrix is
	\begin{align*}
	&B_{\mathrm{tot}}=[B_{\mathrm{ext}},\,B_{\mathrm{site}}], ~~
	J_{\mathrm{tot}}=\operatorname{blkdiag}(J_m,J_n),
\\
	&D_{\mathrm{tot}}=I_{2(m+n)},
	\end{align*}
	where \(B_{\mathrm{ext}}\in\mathbb{R}^{2n\times 2m}\) describes the observable
	external channels and \(B_{\mathrm{site}}\in\mathbb{R}^{2n\times 2n}\) provides
	site-wise hidden damping. The external channels are attached at the endpoints
	of the chain, together with $m-2$ additional evenly distributed sites when
	\(m>2\).

	With an attachment matrix \(S_{\mathrm{ext}}\in\mathbb{R}^{n\times m}\),
	we set
	\[
	B_{\mathrm{ext}}
	=
	\bigl(S_{\mathrm{ext}}\operatorname{diag}(\sqrt{\kappa_{\mathrm{ch}}})\bigr)\otimes I_2,
~
	B_{\mathrm{site}}
	=
	\operatorname{diag}(\sqrt{\kappa_{\mathrm{site}}})\otimes I_2,
	\]
	where the meaning of\(\kappa_{\mathrm{ch}}\) and \(\kappa_{\mathrm{site}}\) is the same as that in Section~\ref{sec:lowchannel_detailed}.
	The full-port PR realization is then
	\begin{equation}
		\begin{aligned}
			A &= J_n R + \tfrac12 B_{\mathrm{tot}} J_{\mathrm{tot}} B_{\mathrm{tot}}^\top J_n,\\
			C_{\mathrm{tot}} &= J_{\mathrm{tot}} B_{\mathrm{tot}}^\top J_n,\\
			D_{\mathrm{tot}} &= I_{2(m+n)}.
		\end{aligned}
		\label{eq:bkc_fullport_template_compact}
	\end{equation}
	The external transfer matrix used for reduction is obtained by restricting to
	the external channels
	\[
	B=B_{\mathrm{ext}},
	\qquad
	C=C_{\mathrm{ext}},
	\qquad
	D=I_{2m},
	\]
	where \(C_{\mathrm{ext}}\) denotes the first \(2m\) rows of \(C_{\mathrm{tot}}\).

	We consider homogeneous and heterogeneous deterministic configurations. In the
	homogeneous case, all sites share the same on-site frequency and local damping,
	whereas in the heterogeneous case both \(\omega_j\) and
	\(\kappa_{\mathrm{site},j}\) vary with \(j\). The latter produces a more
	challenging benchmark with weaker modal uniformity and slower spectral decay.

\begin{figure}[H]
		\centering
		\includegraphics[width=0.48\linewidth]{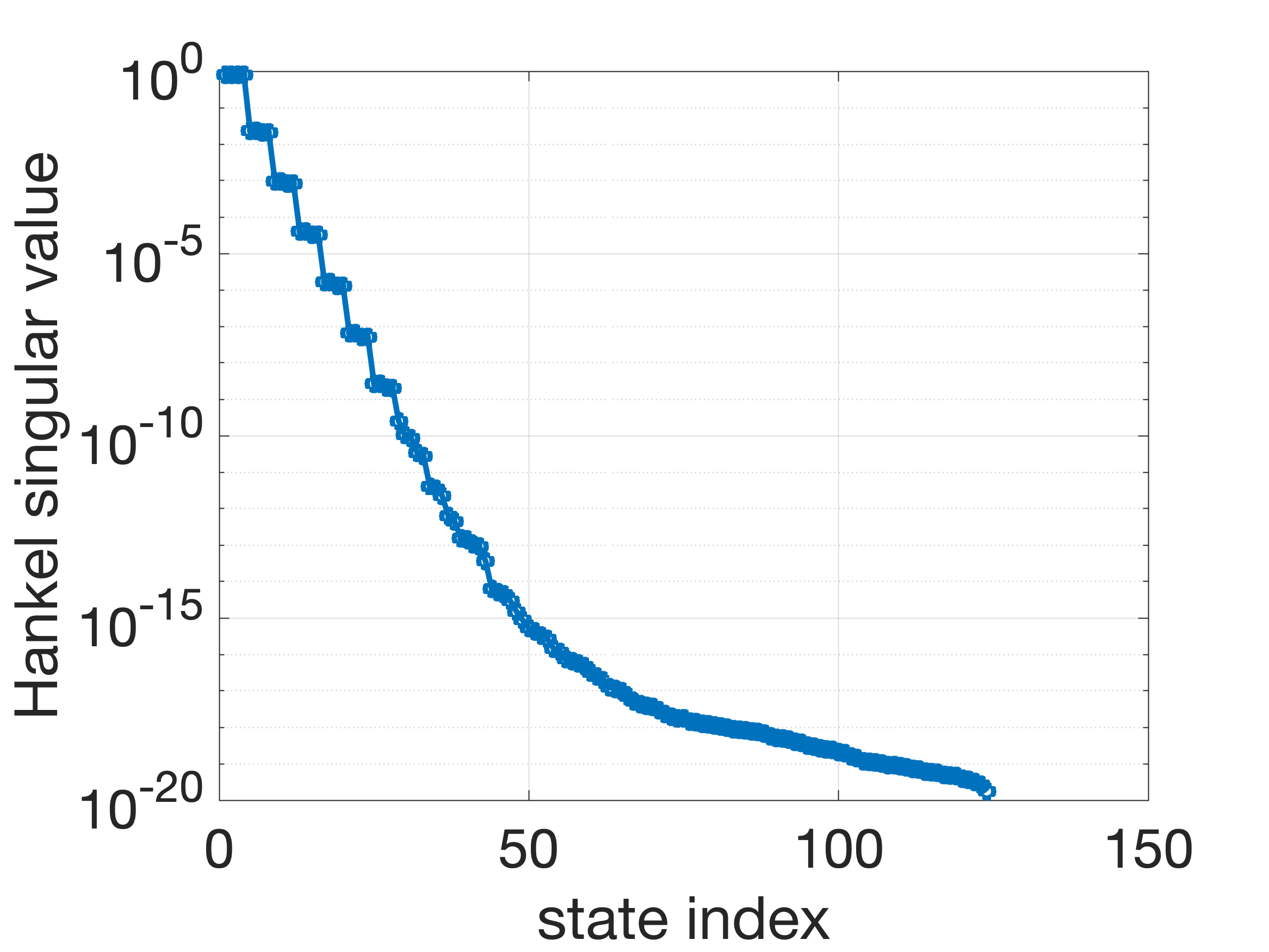}
		\includegraphics[width=0.48\linewidth]{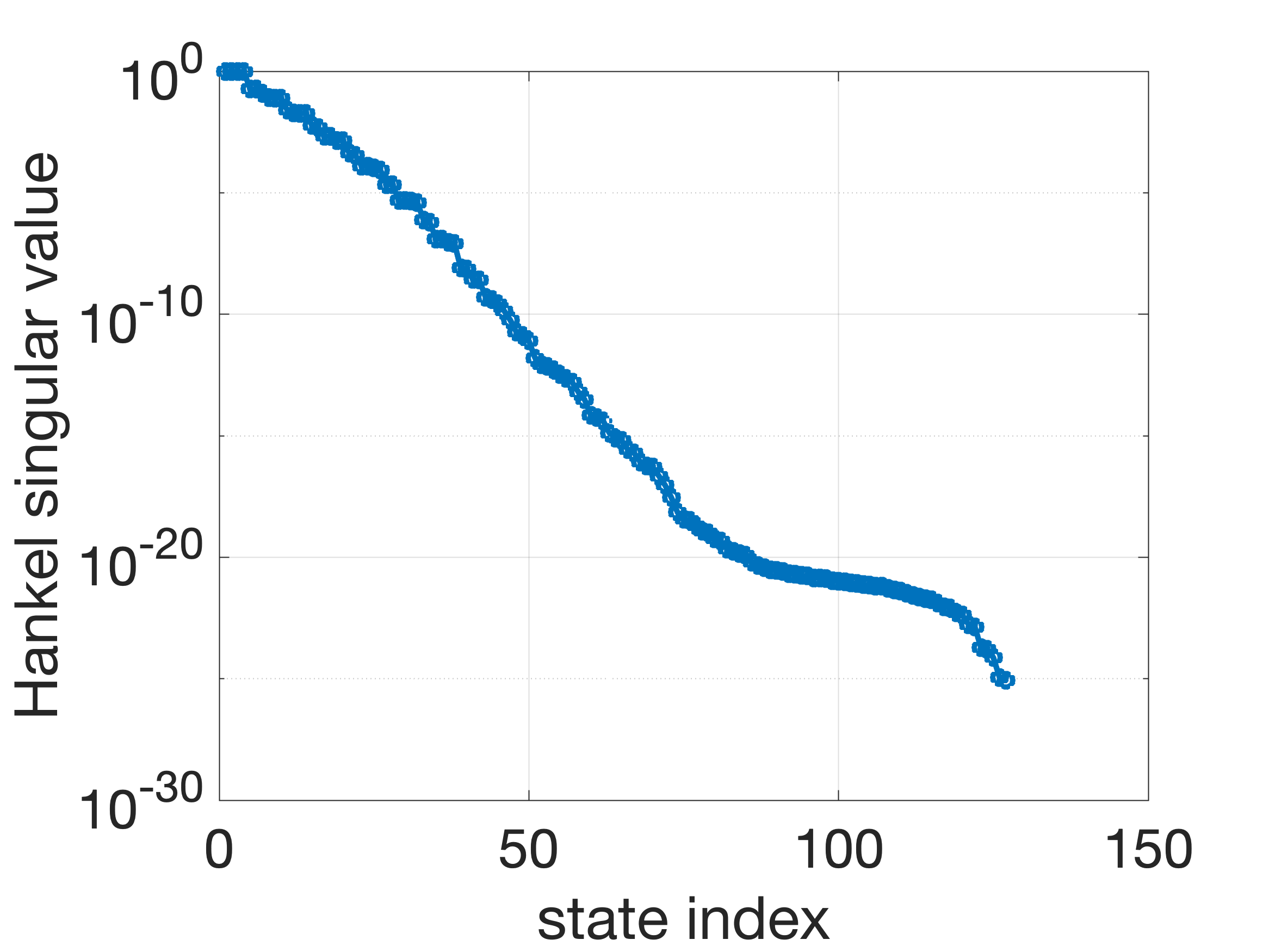}
		\caption{Hankel singular values for the BKC benchmark; homogeneous (left) and heterogeneous (right).}
		\label{fig:hsv_bkc}
	\end{figure}

\begin{figure}[H]
		\centering
		\includegraphics[width=0.48\linewidth]{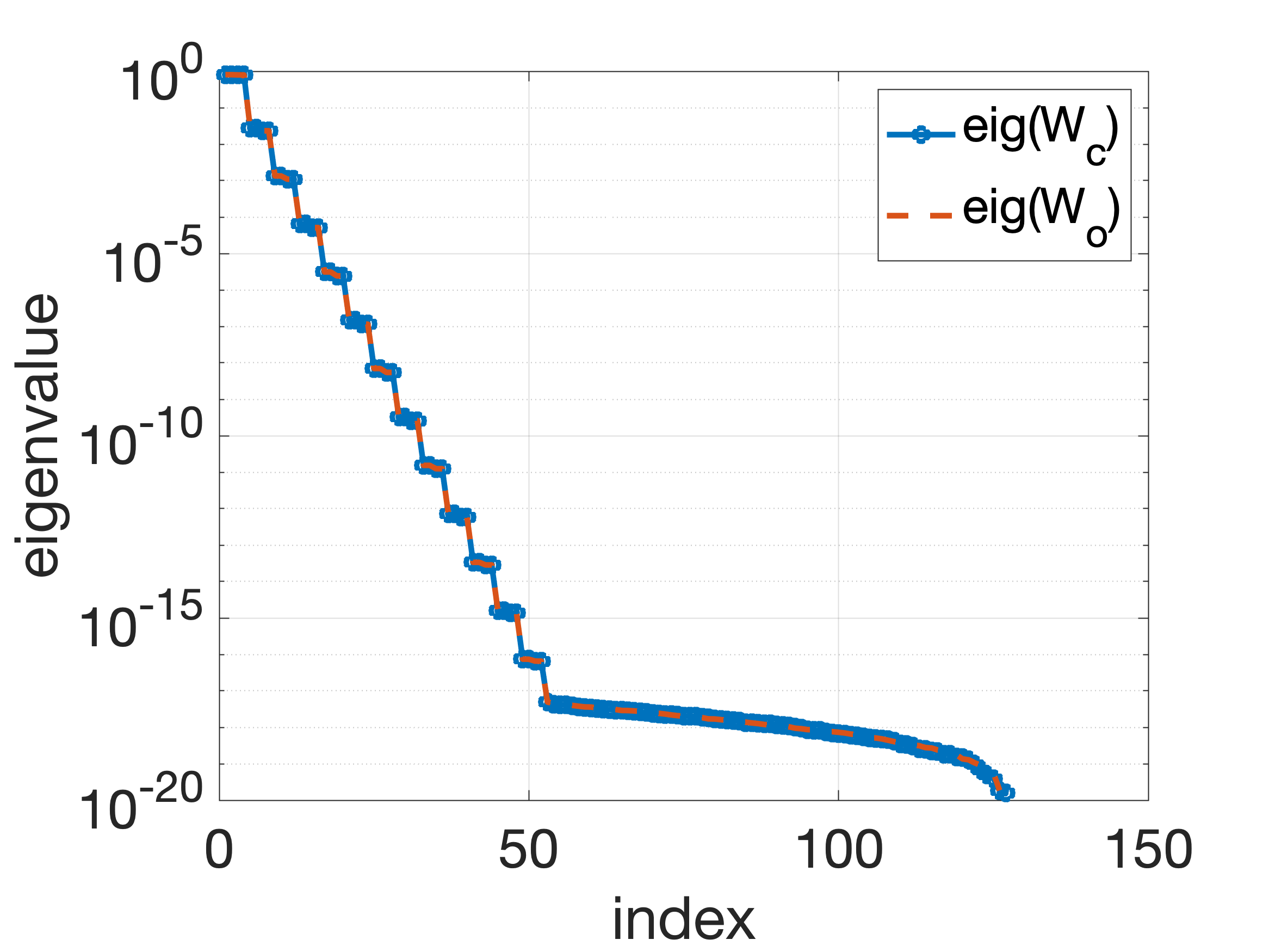}
		\includegraphics[width=0.48\linewidth]{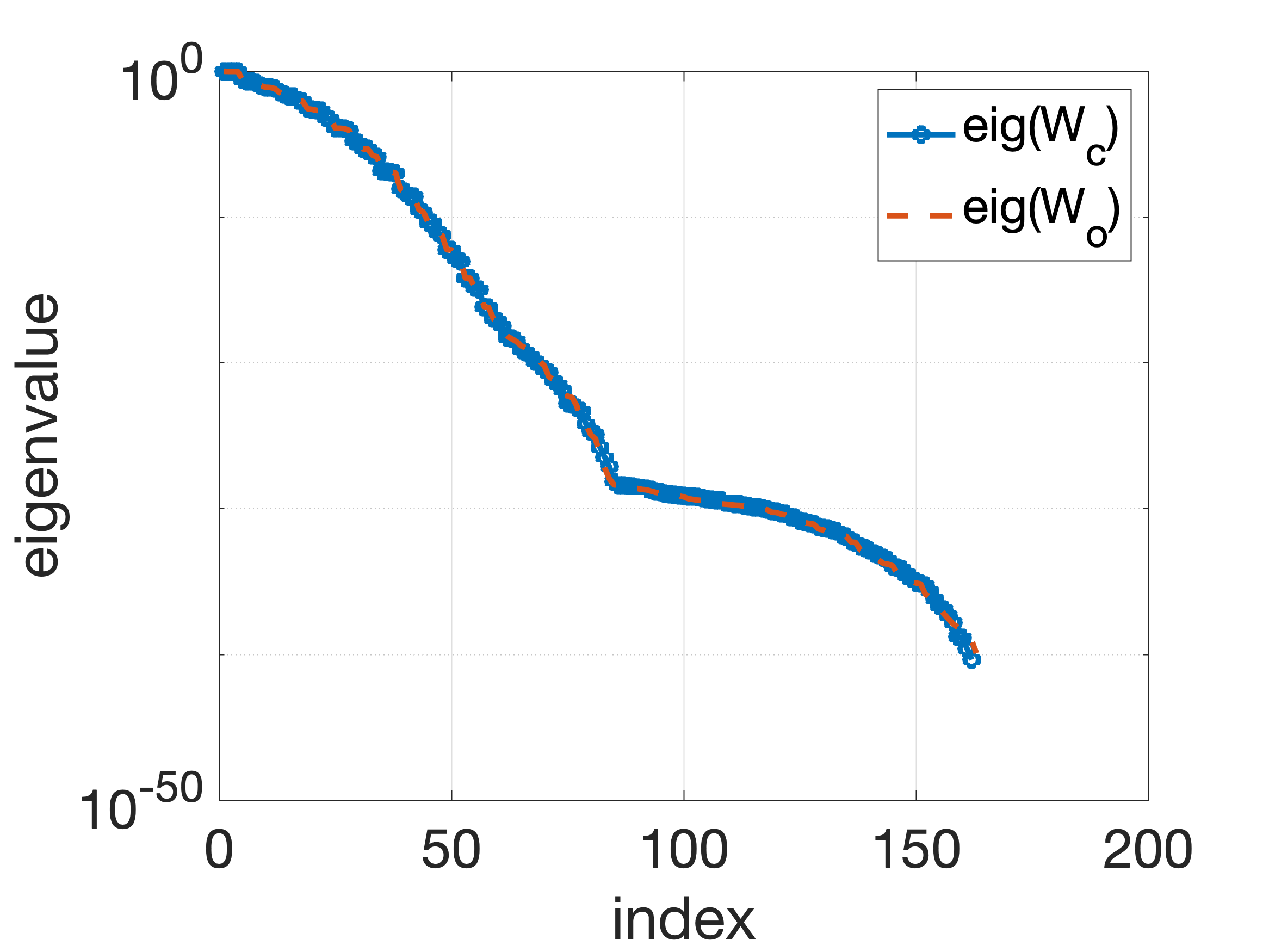}
		\caption{Gramian spectra for the BKC benchmark; homogeneous (left) and heterogeneous (right).}
		\label{fig:gram_bkc}
	\end{figure}

	We first report results for \((n,m,r)=(100,2,10)\). The homogeneous case has a
	larger stability margin,
	\[
	\max \Re(\lambda(A)) \approx -1.25\times 10^{-1},
	\]
	while the heterogeneous case is much less damped,
	\[
	\max \Re(\lambda(A)) \approx -1.5\times 10^{-2}.
	\]
	This difference is reflected in the Hankel singular values: in the homogeneous
	configuration they decay rapidly, with
	\[
	\sigma_{10} \approx 9.6\times 10^{-4},
	\]
	whereas in the heterogeneous case the decay is substantially slower,
	\[
	\sigma_{10} \approx 7.9\times 10^{-2},
	\]
	indicating a significantly larger effective dynamical rank; see
	Figure~\ref{fig:hsv_bkc}. The controllability and observability Gramian spectra
	remain nearly indistinguishable in both cases, again reflecting the PR-induced
	duality, but the heterogeneous benchmark exhibits a slower and more diffuse
	spectral decay; see Figure~\ref{fig:gram_bkc}.

	Q-IRKA is applied to the external-port transfer
	system $(A,B_{\mathrm{ext}},C_{\mathrm{ext}},D_{\mathrm{ext}})$,
	because the approximation objective concerns only the observable external
	input-output map. The hidden site-wise damping channels are not removed from
	the dynamics, that is, their effect is already incorporated into the drift matrix
	\[
	A = J_n R + \tfrac12 B_{\mathrm{tot}} J_{\mathrm{tot}} B_{\mathrm{tot}}^\top J_n,
	\]
	through the full-port physically realizable dilation. Thus the Krylov space is
	constructed from the externally relevant input directions, while the internal
	dissipation induced by the hidden channels remains fully present through the
	resolvent $(sI-A)^{-1}$. \\
	After the projection basis has been computed, the same symplectic projection is
	applied to the full-port
	realization $(A,B_{\mathrm{tot}},C_{\mathrm{tot}},D_{\mathrm{tot}})$,
	so that physical realizability can be verified on the reduced full-port model.
	In this way, the reduced-order model is targeted to the external transfer matrix,
	whereas PR is checked at the level of the full dilation.

\begin{figure}[H]
		\centering
		\includegraphics[width=0.48\linewidth]{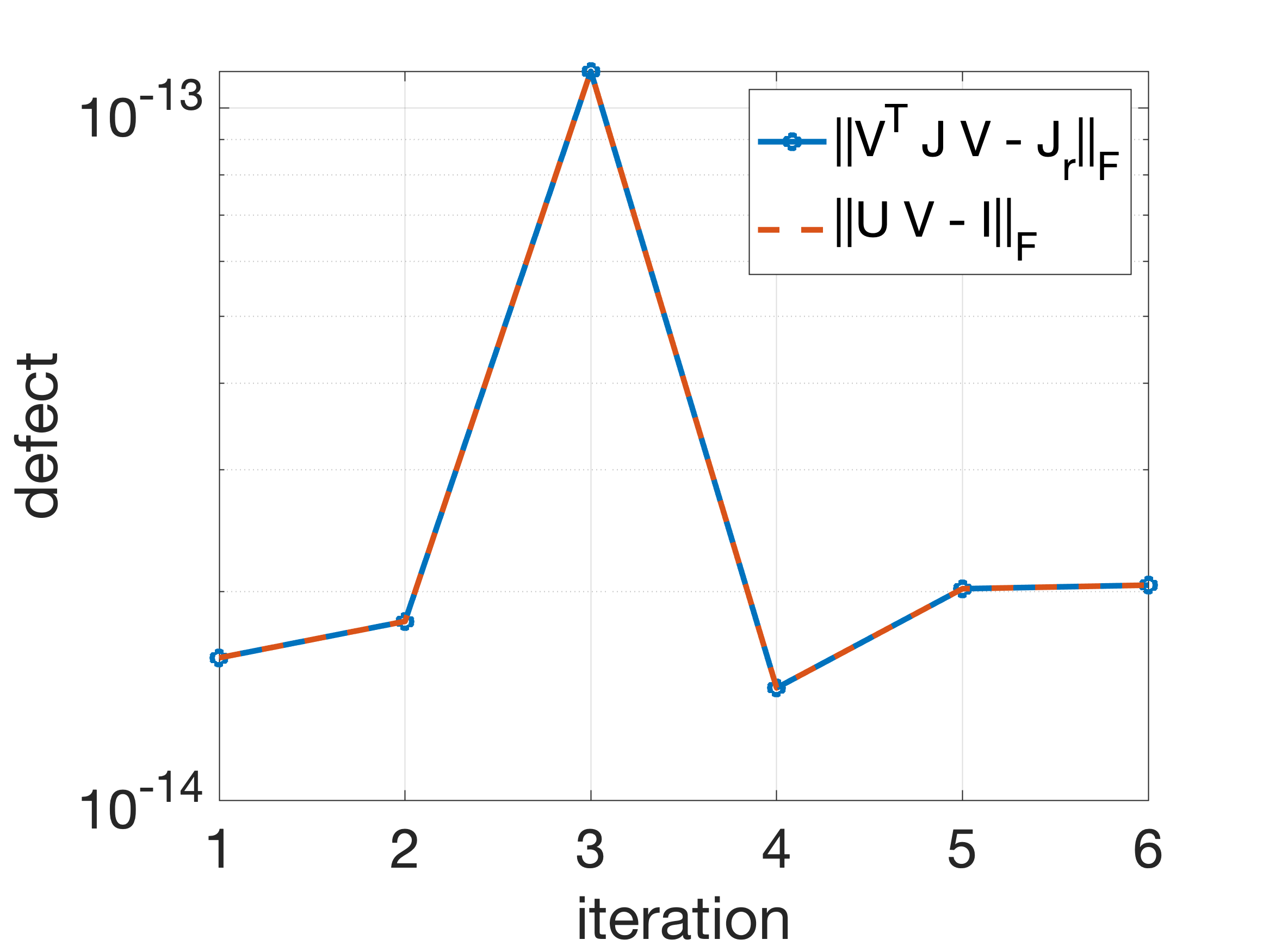}
		\includegraphics[width=0.48\linewidth]{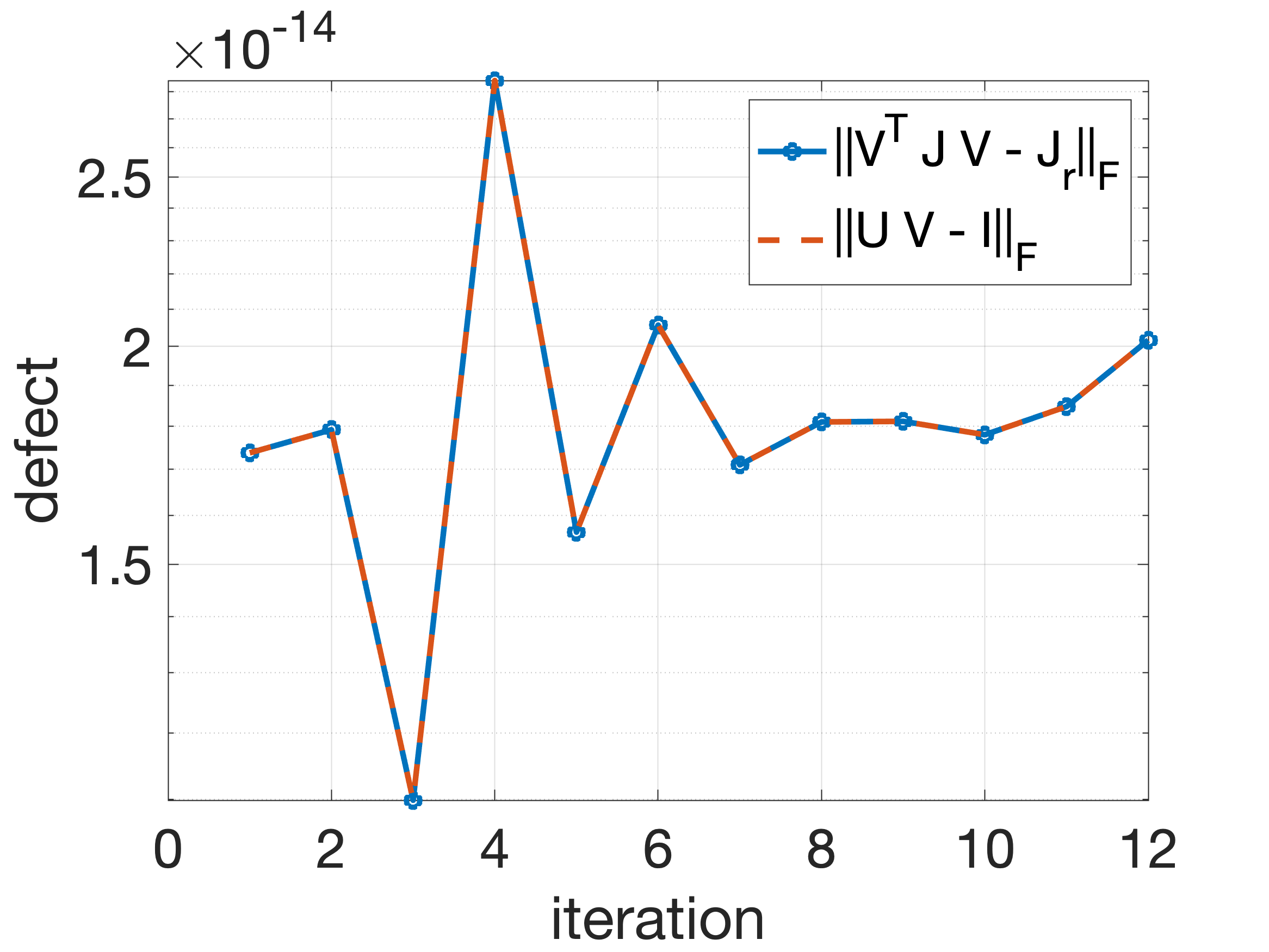}
		\caption{Symplectic and left-inverse defects for the BKC benchmark; homogeneous (left) and heterogeneous (right).}
		\label{fig:symp_bkc}
	\end{figure}

\begin{figure}[H]
		\centering
		\includegraphics[width=0.48\linewidth]{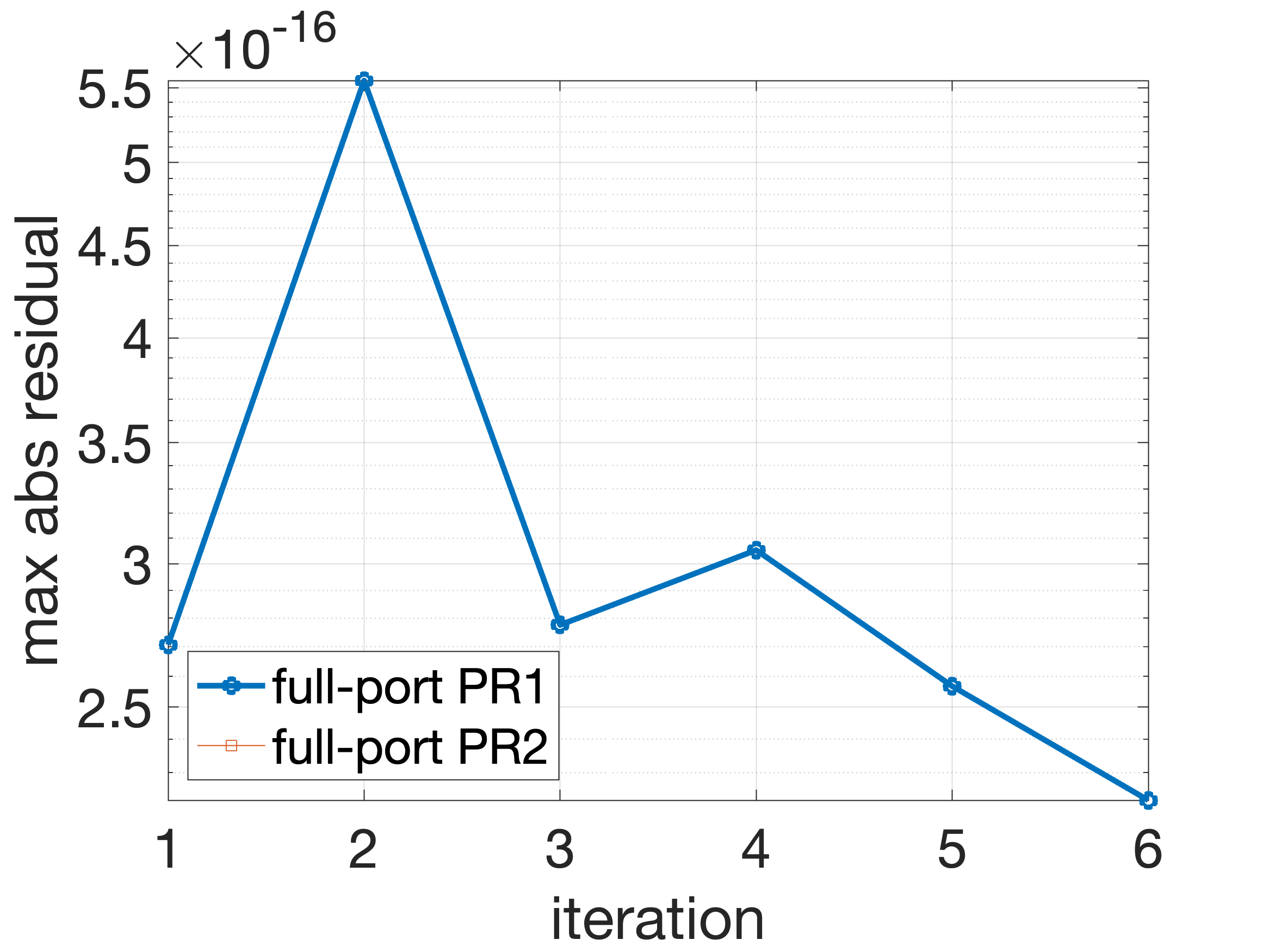}
		\includegraphics[width=0.48\linewidth]{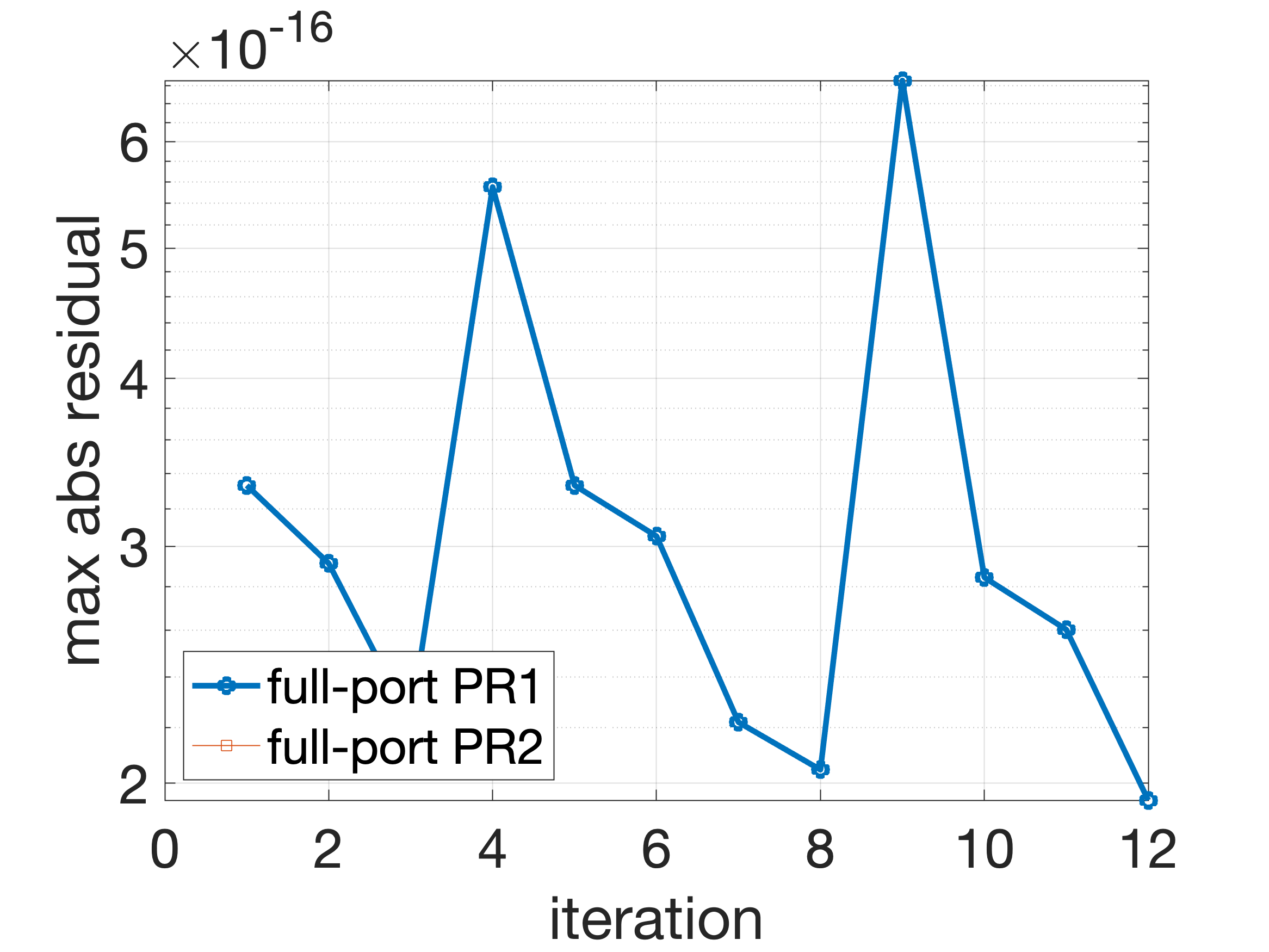}
		\caption{Reduced full-port PR residuals for the BKC benchmark; homogeneous (left) and heterogeneous (right).}
		\label{fig:pr_bkc}
	\end{figure}

	In both homogeneous and heterogeneous configurations, the structural diagnostics remain at machine precision,
	\[
	\Delta_{\mathrm{symp}} \approx 10^{-14},
	\qquad
	\Delta_{\mathrm{PR},j} \approx 10^{-16},
	\]
	showing that symplecticity and physical realizability are preserved robustly
	throughout the iteration; see Figures~\ref{fig:symp_bkc} and \ref{fig:pr_bkc}.

\begin{figure}[H]
		\centering
		\includegraphics[width=0.48\linewidth]{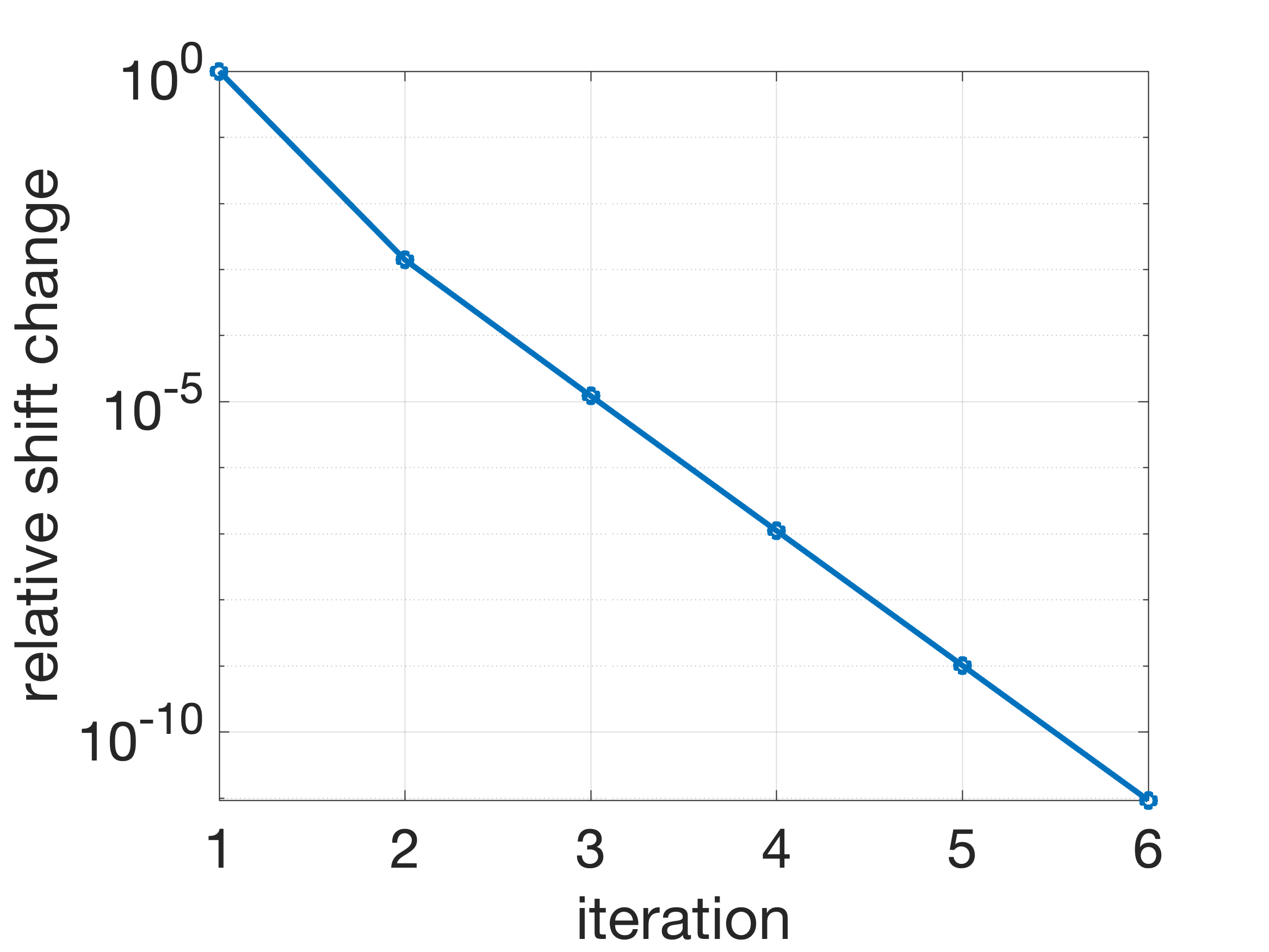}
		\includegraphics[width=0.48\linewidth]{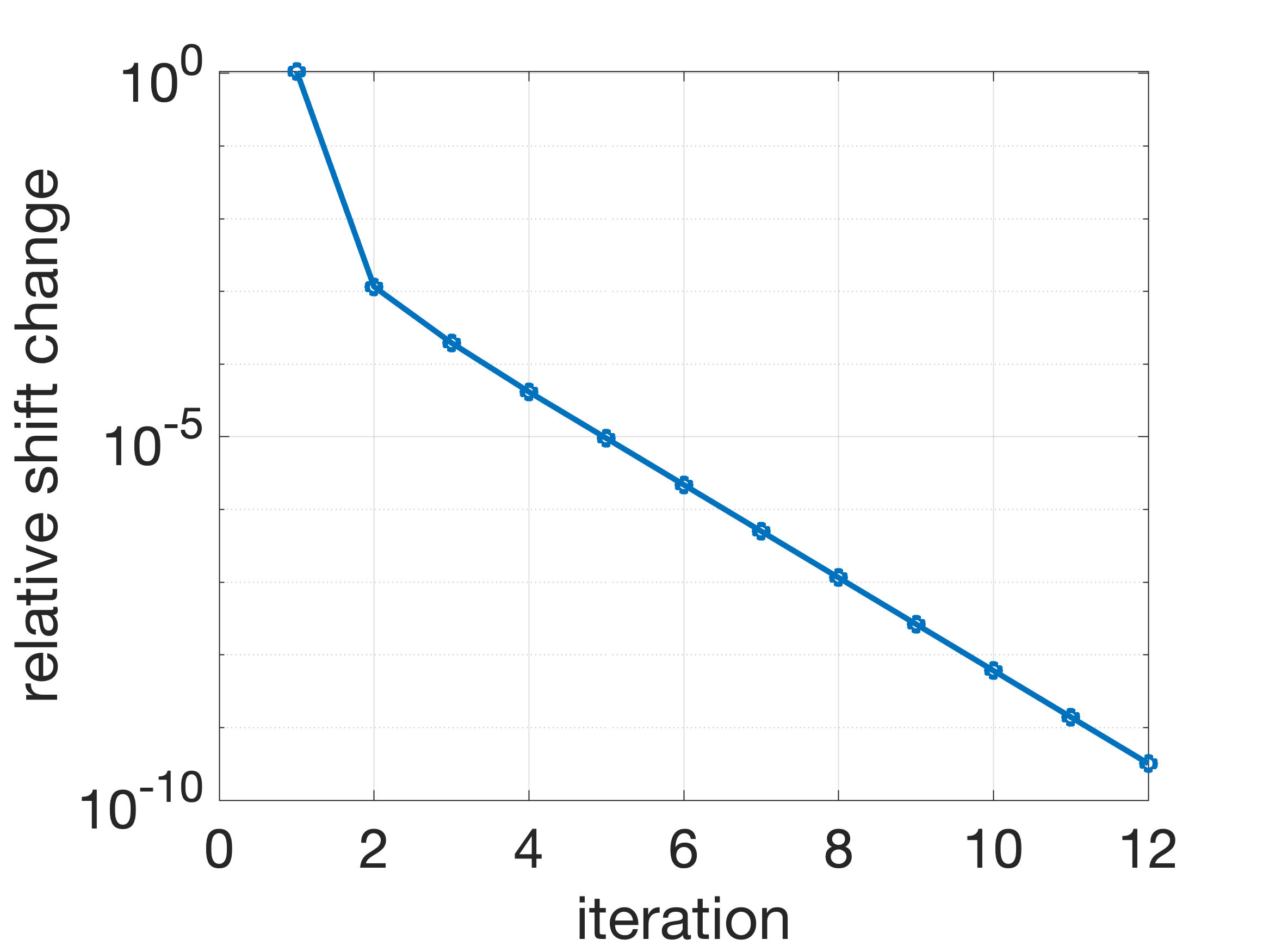}
		\caption{Shift convergence for the BKC benchmark; homogeneous (left) and heterogeneous (right).}
		\label{fig:shift_bkc}
	\end{figure}

	The main algorithmic difference between the two configurations appears in the
	shift iteration. In the homogeneous case, the relative shift change decreases
	rapidly and essentially monotonically, reaching the stopping tolerance within
	six iterations. In the heterogeneous case, convergence is slower and mildly
	non-monotone, although still stable; see Figure~\ref{fig:shift_bkc}. Thus,
	for transport-dominated heterogeneous PR systems, structure preservation and
	fixed-point speed are clearly distinct issues.

	The achieved \(\mathcal H_2\) errors are consistent with this spectral picture.
	In the homogeneous case, the reduced-order model with \(r=10\) is highly accurate,
	with relative error of order \(10^{-5}\). In the heterogeneous case, the error
	is larger, of order \(10^{-3}\), but still moderate relative to the slower
	Hankel decay. Hence, even when the recursive pole update converges more slowly,
	the resulting reduced-order model remains structurally exact and captures the dominant
	external input-output behaviour well.

	To assess complexity, we consider several representative triples
	\((n,m,r)\); see Table~\ref{tab:bkc_complexity}.

	\begin{table*}[t]
		\centering
		\caption{Runtime and relative \(\mathcal H_2\) error for the BKC benchmark.}
		\begin{tabular}{c c c | c c | c c}
			\hline
			$n$ & $m$ & $r$
			& $T_{\mathrm{hom}}$ & $E_{\mathrm{hom}}$
			& $T_{\mathrm{het}}$ & $E_{\mathrm{het}}$ \\
			\hline
			100 & 2 & 10 & 0.518219 & 2.373175e-05 & 1.072527 & 6.840790e-03 \\
			200 & 2 & 10 & 1.767359 & 2.373175e-05 & 4.586289 & 1.346754e-02 \\
			100 & 4 & 10 & 0.438276 & 2.256781e-05 & 1.001470 & 8.391476e-03 \\
			200 & 4 & 10 & 1.354255 & 2.256782e-05 & 4.408679 & 1.282019e-02 \\
			100 & 4 & 20 & 7.101629 & 1.744581e-08 & 1.297693 & 2.276489e-04 \\
			200 & 4 & 20 & 7.578037 & 3.955829e-09 & 11.029683 & 2.927864e-03 \\
			\hline
		\end{tabular}
		\label{tab:bkc_complexity}
	\end{table*}

	Table~\ref{tab:bkc_complexity} may be read by varying one parameter at a time.
	For fixed \(m\) and \(r\), the runtime increases with \(n\); for fixed \(n\)
	and \(m\), it increases more markedly with \(r\). The heterogeneous cases also
	tend to exhibit larger errors and often higher computational cost, reflecting
	their slower spectral decay.
	The computational cost is driven mainly by \(n\) and \(r\), while the
	dependence on \(m\) remains comparatively mild in the tested range. \\
	The approximation error follows the same pattern as in the spectral diagnostics:
	it remains small and nearly independent of \(n\) in the homogeneous case, while
	it increases moderately with \(n\) in the heterogeneous case. Increasing \(r\)
	improves accuracy in both configurations.

	Overall, the BKC benchmark confirms that Q-IRKA preserves symplecticity and
	physical realizability robustly even on transport-dominated systems with
	nontrivial quadrature geometry, while also showing that heterogeneity can
	substantially slow the recursive pole-mirroring iteration without destroying
	reduction quality.

	\section{Conclusion}

	A symplectic interpolatory framework for \(\mathcal{H}_2\) model reduction of
	high-dimensional linear quantum systems was developed under the constraint of
	physical realizability. On this basis, a structure-preserving iterative rational
	Krylov method, referred to as Q-IRKA, was formulated by combining recursive
	pole-based shift updates with symplectic Petrov-Galerkin projection.

	It was shown that physical realizability and symplectic structure were preserved
	by construction throughout the iteration and were maintained numerically to
	machine precision. The reported experiments on low-channel oscillator chains and
	on bosonic Kitaev-chain-inspired benchmarks showed that the reduction quality
	depended strongly on dissipation geometry, heterogeneity, and channel
	configuration, while the computational cost was governed mainly by the system 
	dimension and the reduced order.

	Overall, it was demonstrated that accurate and scalable \(\mathcal{H}_2\) model
	reduction for large-scale linear quantum systems could be achieved without
	sacrificing the underlying physical structure.

The proposed framework also extends naturally to non-square linear quantum
systems, which will be investigated in future work.

	\section{Reproducibility and code}
	The MATLAB scripts and configuration files will be made
	available at\\ \url{https://github.com/alfioborzi}.

\end{document}